\documentclass[11pt]{article}
\usepackage{latexsym}
 \usepackage[dvips]{graphics}
\usepackage{amsmath}  
\usepackage[dvips]{graphics}
\usepackage{latexsym,psfig}
\usepackage{amscd}     \usepackage{amsxtra}
\usepackage{upref}     \usepackage{amsthm}
\usepackage{amssymb}   
\usepackage{amsfonts}
\begin{document}

\title{{\bf  Electromagnetic field objects in terms
of \\ Balance of Geometric flows.}}

\author{{\bf Stoil Donev}\footnote{e-mail:sdonev@inrne.bas.bg},
{\bf Maria Tashkova}, \\
Institute for Nuclear
Research and Nuclear Energy,\\ Bulg.Acad.Sci., 1784 Sofia,
blvd.Tzarigradsko chaussee 72, Bulgaria\\}
\date{}
\maketitle

\begin{abstract}
This paper reviews our physical motivation for choosing appropriate formal
presentation of electromagnetic field objects (EMFO). Our view is based on the
understanding that EMFO are spatially finite entities carrying internal
dynamical structure, so, their available integral time stability should be
represented by appropriate adaptation of their internal dynamical structure to
corresponding local stress-energy-momentum balance relations with other
physical objects. This adaptation process has two aspects: internal and
external. Clearly, finding adequate internal dynamical structure giving
appropriate integral characteristics of the object, will bring also appropriate
behavior of EMFO as a whole. Therefore, the internal local
stress-energy-momentum balance among the subsystems of EMFO should formally be
presented by appropriately defined tensor-field quantities, which are meant to
suggest a dinamical understanding of the abilities of EMFO to successfully, or
not successfully, communicate with all the rest physical world.

\end{abstract}

{\it keywords}: extended electrodynamics, electromagnetic field objects.

\section{Introduction}

Modern theoretical view on classical fields accepts that
time dependent and space propagating electromagnetic fields are flows
of time stable physical entities which have been called in
the early 20th century {\it photons}. Since appropriate in this respect
nonlinearizations of Maxwell vacuum equations are still missing
[1],[2],[3],[4],[5],[6],[7],[8],[9],[10],[11],[12],[13],[14],[15],
and the seriously developed quantum theory also does not give
appropriate, from our viewpoint, description of time stable entities of
electromagnetic field nature, we decided to look back to the rudiments of the
electromagnetic theory trying to reconsider its assumptions in order to come to
equations giving appropriate solutions, in particular, solutions, demonstrating
internal dynamical structure, having finite spatial carrier at every moment of
their existence, and space-propagating as a whole, keeping, of course, their
physical identity and recognizability.

According to our view, in presence of appropriate environment, the dynamical
equations, describing locally, i.e., around every point
inside the spatial carrier, the evolution of the object, may come from giving
an explicit form of the quantities, controlling the local internal and external
exchange processes, in other words, the equations must express corresponding
{\it local balance relations}.

We note that the properties {\it spatial finiteness} and {\it internal dynamical
structure} we consider as very essential ones. So, the classical material
points and the infinite classical fields (e.g.  plane waves) should not be
considered as models of physical objects since the former have no structure and
cannot be destroyed at all, and the latter carry infinite energy, so they
cannot be finite-time created. Therefore, the Born-Infeld "principle of
finiteness" [2] stating that {\it "a satisfactory theory should avoid letting
physical quantities become infinite"} we strengthen as follows:

\vskip 0.3cm
{\bf All real physical objects are spatially finite entities and NO infinite
values of the physical quantities carried by them should be allowed}.

\vskip 0.2cm
Clearly, together with the purely qualitative features, physical objects carry
important quantitatively described physical properties, and any external
interaction may be considered as an exchange of the corresponding quantities
provided both, the object and the corresponding environment, carry them. Hence,
the more universal is a physical quantity the more useful for us it is, and
this moment determines the exclusively important role of
stress-energy-momentum, which modern physics considers as the most universal
one, so we may assume that:
\vskip 0.3 cm
{\bf Propagating electromagnetic field objects necessarily carry
energy-momentum.}
\vskip 0.3cm
Further in the paper we shall follow the rule:
\vskip 0.2cm
{\bf Physical recognizability of time-stable subsystems of a
physical system requires corresponding mathematical recognizability in the
theory}.
\vskip 0.2cm
Assuming that any physical interaction presupposes
dynamical flows of some physical quantities among the subsystems of the
physical system considered, the field nature of the objects  suggests {\it
local nature} of these flows, so, {\bf every continuous subsystem is supposed to
be able to build CORRESPONDING LOCAL INSTRUMENTS, realizing explicitly the
flows}. In static cases these flows reduce, of course, to {\it stress}.
Formally this means: \begin{center} \hfill\fbox{
    \begin{minipage}{0.97\textwidth}
\vskip 0.3cm
	1. We must have a mathematical field object $\mathcal{A}$ representing
the system as a whole.
\vskip 0.3cm
	2. The supposed existence of recognizable and mutually
interacting subsystems $(A_1,A_2,...)$ of $\mathcal{A}$ leads to the assumption
for {\it real} but {\it admissible}, i.e., not leading to annihilation, changes
of the subsystems, so, such changes should be formally represented by tensor
objects.
 \vskip 0.3cm
	3. The local flow manifestation of the admissible real changes suggests
to make use of appropriate combination of tensor objects, corresponding tensor
co-objects, and appropriate invariant differential operators.
 \vskip 0.3cm
	4. Every coupling inside this combination shall distinguish
existing partnership, i.e., interaction, among the subsystems, so, all such
couplings should be duly respected when the system's time-stability and spatial
propagation are to be understood.
\vskip 0.5cm
\end{minipage}}
 \hfill \end{center}

In order to make our view more easily and rightly apprehended we begin with the
strongly idealized example of a {\it static} classical field object, if for {\it
mathematical images} of the {\it physical constituents}, further called {\it
formal constituents}, of the object are chosen vector fields on the
traditional classical space $\mathbb{R}^3$.

\section{Maxwell stress tensors}

Every vector field, defined on an arbitrary manifold $M$, generates 1-parameter
family $\varphi_t$ of (local in general) diffeomorphisms of $M$. Therefore,
having defined a vector field $X$ on $M$, we can consider for each
$t\in\mathbb{R}$ the corresponding diffeomorphic image $\varphi_t(U)$ of any
region $U\subset M$. Hence, interpreting the external parameter $t$ as time,
which is NOT obligatory, vector fields may be formally tested as formal images
of the dynamical constituents of some spatially finite real field objects.

Let now $X$ be a vector field on the euclidean space $(\mathbb{R}^3,g)$, where
$g$ is the euclidean metric in $T\mathbb{R}^3$, having in the canonical global
coordinates $(x^1,x^2,x^3=x,y,z)$ components $g_{11}=g_{22}=g_{33}=1$, and
$g_{12}=g_{13}=g_{23}=0$. The induced euclidean metric in $T^*\mathbb{R}^3$ has
in the dual bases the same components and will be denoted further by the same
letter $g$. The corresponding isomorphisms between the tangent and cotangent
spaces and their tensor, exterior and symmetric products will be denoted by
the same simbol $\tilde{g}$, so (summation on the repeating indecies is assumed)
$$
\tilde{g}\left(\frac{\partial}{\partial x^i}\right)=
g^{ik}\left(\frac{\partial}{\partial x^k}\right)=dx^i,
\ \ (\tilde{g})^{-1}(dx^i)=\frac{\partial}{\partial x^i} \ \  \cdots
$$

Having a co-vector field, i.e. 1-form $\alpha$, or
another vector field  $Y$ on $\mathbb{R}^3$, we can form the {\it flow of} $X$
{\it across} $\alpha$, or across the $\tilde{g}$-coobject $\tilde{g}(Y)$ of $Y$:
$$
i_{X}\alpha=\langle\alpha,X\rangle=\alpha_1X^1+\alpha_2X^2+\alpha_3X^3,
$$
$$
i_{X}\tilde{g}(Y)=\langle\tilde{g}(Y),X\rangle=
g(X,Y)\equiv X.Y=g_{ij}X^iY^j=X_iY^i=X_1Y^1+X_2Y^2+X_3Y^3.
$$
This flow of $X$ is an invariant entity, so to its admissible
and appropriate changes should be paid due respect. According to classical
vector analysis on $\mathbb{R}^3$ [16] for the differential of the function
$g(X,Y)$ we can write
$$
(\tilde{g})^{-1}[\mathbf{d}g(X,Y)]=(X.\nabla)Y+(Y.\nabla)X+X\times\it{curl}(Y)+
Y\times\it{curl}\,(X),
$$
where in our coordinates
$$
X.\nabla=\nabla_{X}=X^i\frac{\partial}{\partial x^i}, \ \ \
(X.\nabla)Y=\nabla_{X}Y=
X^i\frac{\partial Y^j}{\partial x^i}\frac{\partial}{\partial x^j},
$$
"$\times$" denotes the usual vector product, and
$$
(\it{curl}X)^i=\left(\frac{\partial X^3}{\partial x^2}-\frac{\partial
X^2}{\partial x^3}, \ \frac{\partial X^1}{\partial x^3}-\frac{\partial
X^3}{\partial x^1}, \ \frac{\partial X^2}{\partial x^1}-
\frac{\partial X^1}{\partial x^2}\right)\cdot
$$
The Hodge $*_{g}$-operator acts in these coordinates as follows:
$$
*dx=dy\wedge dz, \ \ *dy=-dx\wedge dz, \ \ *dz=dx\wedge dy,
$$
$$
*(dx\wedge dy)=dz, \ \ *(dx\wedge dz)=-dy, \ \ *(dy\wedge dz)=dx,
$$
$$
*(dx\wedge dy\wedge dz)=1, \ \ \ *1=dx\wedge dy\wedge dz .
$$
{\bf Corollary}. The following relation holds ($\mathbf{d}$ denotes the
exterior derivative):
$$
\it{curl}(X)=(\tilde{g})^{-1}\,
*\,\mathbf{d}\,\tilde{g}(X), \ \ \text{or} \ \
\tilde{g}(\it{curl}X)=*\,\mathbf{d}\,\tilde{g}(X).
$$
Assume now that in the above expression for $\mathbf{d}g(X,Y)$ we put $X=Y$,
i.e., we consider the invariant local change of the flow of $X$ across its
proper coobject $\tilde{g}(X)$. We obtain
 $$
\frac12\mathbf{d}g(X,X)=\frac12\mathbf{d}(X^2)=
\tilde{g}(X\times\it{curl}X+(X.\nabla)X)=
\tilde{g}(X\times\it{curl}X+\nabla_XX).
$$
In components, the last term on the right reads
$$
(\nabla_XX)^j=X^i\nabla_i X^j=\nabla_i(X^iX^j)-X^j\nabla_iX^i=
\nabla_i(X^iX^j)-X^j\mathrm{div}\,X ,
$$
where (denoting by $L_X$ the Lie derivative along $X$)
$$
\mathrm{div}X=*L_X\left(dx\wedge dy\wedge dz\right)=
*\left(\frac{\partial X^i}{\partial x^i}dx\wedge dy\wedge dz\right)=
\frac{\partial X^i}{\partial x^i}.
$$
Substituting into the preceding relation, replacing
$\mathbf{d}(X^2)$ by $(\nabla_i\delta^i_jX^2)dx^j$, where $\delta^i_j$ is the
unit tensor in $T\mathbb{R}^3$, and making some elementary transformations we
obtain
$$
\nabla_i\left(X^iX^j-\frac12
g^{ij}X^2\right)= \big [(\it{curl}\,X)\times X+X\mathrm{div}\,X\big ]^j.
$$
The symmetric 2-tensor
\begin{equation}
M^{ij}=X^iX^j-\frac12\,g^{ij}X^2=
\frac12\Big
[X^iX^j+\big(\tilde{g}^{-1}\circ*\tilde{g}(X)\big)^{ik}
\big(*\tilde{g}(X)\big)_{k}\,^{j}\Big ]
\end{equation}
we shall call further Maxwell stress tensor generated by the (arbitrary) vector
field $X\in\mathfrak{X}(\mathbb{R}^3)$. The components $M^i_j$ represent the
generated by the dynamical nature of $X$ local stresses, and the local stress
energy is represented in terms of $tr(M^i_j)=M^i_i$. Momentum is missing since
time is a missing dimension. {\it The appropriate changes of $M^i_j$, given in
this idealised case by $(curl\,X\times X)$ and $(X\,div\,X)$, could be
considered as possible local instruments in terms of which, in presense of more
constituents and recognizable subsystems, corresponding
balance relations to be written down}.

We specially note that, formally, $M^{ij}$ may be represented as sum of the
stresses carried by $X$ and by the 2-vector
$\tilde{g}^{-1}*\tilde{g}(X)=\tilde{g}^{-1}i_{X}(dx\wedge dy\wedge dz)$.
Hence, since $\tilde{g}^{-1}*\tilde{g}(X)$ is uniquely determined by $X$ and
$g$, we may consider an idealised physical object,
built of two constituents
$X$ and $\tilde{g}^{-1}*\tilde{g}(X)$.
It should be noted however that {\it interacting stress} between the two
costituents is missing: $M^{ij}$ is sum of the stresses carried by $X$ and
$\tilde{g}^{-1}*\tilde{g}(X)$.

Clearly, when we raise and lower indices
in canonical coordinates with $\tilde{g}$ we shall have the following component
relations:
$$
M_{ij}=M_i^j=M^{ij}
$$ which does not mean, of course, that we equalize
quantities being elements of different linear spaces.

We note now some formal relations.

First, the easily verified relation between the vector product "$\times$"
and the wedge product in the space of 1-forms on $\mathbb{R}^3$:
$$
X\times Y=(\tilde{g})^{-1}\,(*\,(\tilde{g}(X)\wedge\tilde{g}(Y)))
$$
$$
=(\tilde{g})^{-1}\,\circ i(X\wedge Y)(dx\wedge dy\wedge dz), \ \
X,Y\in\mathfrak{X}(\mathbb{R}^3) .
$$

We are going to consider now the differential flow nature of
$\nabla_iM^i_jdx^j$.

	{\bf Proposition}. If $\alpha=\tilde{g}(X)$ then the following relation
holds ($i(X)\mathbf{d}\alpha$ means $X^i\mathbf{d}\alpha_{ij}dx^j$):
$$
\tilde{g}(\it{curl}\,X\times X)=i(X)\mathbf{d}\alpha=
-*(\alpha\wedge *\mathbf{d}\alpha) .
$$
{\it Proof}.
$$\tilde{g}(\it{curl}\,X\times X)=
\tilde{g}\circ(\tilde{g})^{-1}\,*(\tilde{g}(\it{curl}X)\wedge\tilde{g}(X))=
-*(\alpha\wedge *\mathbf{d}\alpha).
$$
For the component of $i(X)\mathbf{d}\alpha$ before $dx$ we obtain
$$
-X^2\left(\frac{\partial \alpha_2}{\partial x^1}-\frac{\partial
\alpha_1}{\partial x^2}\right)-X^3\left(\frac{\partial \alpha_3}{\partial
x^1}- \frac{\partial \alpha_1}{\partial x^3}\right),
$$
and the same quantity is easily obtained for the component of $[-*(\alpha\wedge
*\mathbf{d}\alpha)]$ before $dx$. The same is true for the components of the
two 1-forms before $dy$ and $dz$. The proposition is proved. Since a 2-form may
be considered as a 2-volume form on $U^{2}\subset\mathbb{R}^3$, we may
interpret the above relation in the sense that,
$\mathbf{d}\alpha=\mathbf{d}\tilde{g}(X)$ is a volume 2-form across which the
vector field $X$ will drag the points of the finite region
$U^2\subset\mathbb{R}^3$.

As for the second term
$X\mathrm{div}X$ of the divergence $\nabla_iM^{ij}$, since
$\mathbf{d}*\alpha=\mathrm{div}X(dx\wedge dy\wedge dz)$, we easily obtain
\begin{equation}
i(\tilde{g}^{-1}(*\alpha))\mathbf{d}*\alpha=(\mathrm{div}X)\alpha.
\end{equation}
Hence, analogically, the 2-vector $\tilde{g}^{-1}(*\alpha)$ will drag the
points of $U^2\subset\mathbb{R}^3$ across the 3-form $\mathbf{d}*\alpha$,
meaning that the 1-form $\alpha$ changes to $(\mathrm{div}X)\alpha$.

We can write now
\begin{equation}
M_{ij}=\frac12[\alpha_i\alpha_j+(*\alpha)_i\,^k(*\alpha)_{kj}], \ \ \
\nabla_iM^i_j=\big[i(X)\mathbf{d}\alpha +
i(\tilde{g}^{-1}(*\alpha))\mathbf{d}*\alpha\big]_j .
\end{equation}
So, a stress balance between the formal constituents $X$
and $\tilde{g}^{-1}(*\alpha)$ is described by
\begin{equation}
i(X)\mathbf{d}\alpha=-i(\tilde{g}^{-1}(*\alpha))\mathbf{d}*\alpha.
\end{equation}

A formal suggestion that comes from the above relations (4) is:
\vskip 0.2cm
{\bf The interior
product of a (multi)vector and a differential form (i.e. the flow of a
(multi)vector field across a differential form) may be considered as
appropriate means, generating quantitative measure of local physical
interaction.}
\vskip 0.2cm

Hence, the naturally isolated two
terms in $\nabla_iM^i_jdx^j$ suggest: any realizable static stress,
that can be associated with the vector field $X$, to be described by
$(\alpha,*\alpha)$, and the recognizable nature of $\alpha$ and $*\alpha$
to be guaranteed by the balance equation $i(X)\mathbf{d}\alpha=-
i(\tilde{g}^{-1}(*\alpha))\mathbf{d}*\alpha$,
or by
$$
i(X)\mathbf{d}\alpha=0, \ \
i(\tilde{g}^{-1}(*\alpha))\mathbf{d}*\alpha=0, \ \ \ \text{i.e.}, \ \ \ \
X\times\it{curl}\,X=0, \ \mathrm{div}\,X=0.
$$

Recalling now how the Lie derivative with respect to (multi)vectors
 acts on 1-forms and 2-forms [17], namely,
$$
L_X\alpha=\mathbf{d}\langle\alpha,X\rangle+i(X)\mathbf{d}\alpha, \ \
\text{i.e.,} \ \ \ \
L_X\alpha-\mathbf{d}\langle\alpha,X\rangle=i(X)\mathbf{d}\alpha,
$$
$$
L_{\bar{*\alpha}}*\alpha=\mathbf{d}\langle*\alpha,\bar{*\alpha}\rangle-
(-1)^{deg{(\bar{*\alpha})}.deg{(\mathbf{d})}}i(\bar{*\alpha})\mathbf{d}*\alpha,
\ \ \text{i.e.} \ \ \
L_{\bar{*\alpha}}*\alpha-\mathbf{d}\langle*\alpha,\bar{*\alpha}\rangle=
-i(\bar{*\alpha})\mathbf{d}*\alpha, \ \
$$
where $deg(\bar{*\alpha})=2, \  deg{(\mathbf{d})}=1$,
we see that the flow of $X$ across the 2-form $\mathbf{d}\alpha$, and the flow
of the 2-vector $\tilde{g}^{-1}(*\alpha)$ across $\mathbf{d}*\alpha$, are given
by the difference between two well defined coordinate free quantities, and this
difference determines when the local change $\mathbf{d}\alpha$
 of $\alpha$ with
respect to $X$, resp. $\mathbf{d}*\alpha$ of $*\alpha$ with respect to
$\tilde{g}^{-1}(*\alpha)$, cannot be represented by
$\mathbf{d}\langle\alpha,X\rangle$,
resp. $\mathbf{d}\langle*\alpha,\tilde{g}^{-1}(*\alpha)\rangle$.

\vskip 0.3cm

We pass now to the case of two formal constituents represented by {\it two}
vector fields. \vskip 0.2cm Let $V$ and $W$ be two vector fields on our
euclidean 3-space. Summing up the corresponding two Maxwell stress tensors
(Sec.1) we obtain the identity:

\setlength\arraycolsep{8pt}
\begin{eqnarray*}
\lefteqn{ \nabla_iM^{ij}_{(V,W)}\equiv \nabla_i\left(V^iV^j+W^iW^j-
g^{ij}\frac{V^2+W^2}{2}\right)={} } \nonumber\\ & & {}=\big
[(\it{curl}\,V)\times V+ V\mathrm{div}\,V+
(\it{curl}\,W)\times W+W\mathrm{div}\,W\big ]^j.
\end{eqnarray*}
Note that the balance in this case {\bf may} look like, for example, as follows:
$$
(\it{curl}\,V)\times V=-W\mathrm{div}\,W, \ \ \
(\it{curl}\,W)\times W=-V\mathrm{div}\,V,
$$
which suggests internal/mutual stress balance between two
subsystems created by two constituents formally described by $V$ and $W$.

Let now $(a(x,y,z),b(x,y,z))$ be two arbitrary functions on $\mathbb{R}^3$. We
consider the transformation
$$
(V,W)\rightarrow (V\,a-W\,b,V\,b+W\,a).
$$

We specially note:

1. The tensor $M_{(V,W)}$ transforms to $(a^2+b^2)M_{(V,W)}$.

2. The transformations
$(V,W)\rightarrow (V\,a-W\,b,V\,b+W\,a)$ do not change the eigen
directions structure of $M^{ij}_{(V,W)}$.

3. If $a=\mathrm{cos}\,\theta, b=\mathrm{sin}\,\theta$, where
$\theta=\theta(x,y,z)$ then the tensor $M_{(V,W)}$ stays invariant:
$$
M_{(V,W)}=M_{(V\mathrm{cos}\,\theta-W\mathrm{sin}\,\theta,
V\mathrm{sin}\,\theta+W\mathrm{cos}\,\theta)}.
$$

The expression inside the parenteses above, denoted by $M^{ij}_{(V,W)}$, looks
formally the same as the introduced by Maxwell tensor
$M^{ij}(\mathbf{E},\mathbf{B})$ from physical considerations concerned with the
electromagnetic stress energy properties of continuous media in presence of
external electromagnetic field $(\mathbf{E},\mathbf{B})$. Formally any vector
$V$, or any couple of vectors $(V,W)$, define such tensor (which we denoted by
$M_V$, or $M_{(V,W)}$), called further {\bf Maxwell stress tensor}. The
term, "stress" in this general mathematical setting could be justified by the
above mentioned dynamical nature of vector fields. It deserves noting here that
the two-vector case  should be expected to satisfy some conditions of
compatability between $V$ and $W$ in order to physically represent some
mutually balanced time stable stress flows.

We emphasize the following moments:
\vskip 0.2cm
	{\bf 1}. The differential identity satisfied by $M_{(V,W)}$
is purely mathematical;

	{\bf 2}. On the two sides of this identity stay well defined
coordinate free quantities;

	{\bf 3}. The tensors $M_{(V,W)}$ do NOT introduce
interaction stress: {\bf the full stress is the sum of the stresses generated by
each one of the constituents} $(V,W)$.
\vskip 0.2cm
Physically, we may say that the corresponding physical medium that occupies the
spatial region $\mathbf{U}_o$ and is parametrized by the points of the
mathematical subregion $U_o\subset\mathbb{R}^3$, is subject to {\it compatible}
and {\it admissible} physical "stresses", and these physical stresses are
quantitatively described by the corresponding physical interpretation of the
tensor $M_{(V,W)}$. Clearly, we could extend the couple $(V,W)$ to more vectors
$(V_1,V_2,...,V_p)$, but then the mentioned invariance properties of
$M_{(V,W)}$ may be lost, or should be appropriately extended.

Finally, note that the stress tensor $M^{ij}$ appears as been subject to the
divergence operator, and if we interpret the components of $M^{ij}$ as physical
stresses, then its divergence acquires, in general, the physical interpretation
of {\it force density}. Of course, in the {\it static} situation as it is given
by the relation considered, no stress propagation is possible, so at every
point the local forces mutually compensate: $\nabla_{i}M^{ij}=0$.

Now, analizing the eigen and other properties of the Maxwell energy tensors,
we try to find some appropriate suggestions.

\section{What the properties of Maxwell stress tensors suggest.}

We consider $M^{ij}(\mathbf{E},\mathbf{B})$ at some point $p\in\mathbb{R}^3$
and assume that in general the vector fields $\mathbf{E}$ and $\mathbf{B}$ are
lineary independent, so $\mathbf{E}\times\mathbf{B}\neq 0$. Let the coordinate
system be chosen such that the coordinate plane $(x,y)$ to coincide with the
plane defined by $\mathbf{E}(p),\mathbf{B}(p)$. In this coordinate system
$\mathbf{E}=(E_1,E_2,0)$ and $\mathbf{B}=(B_1,B_2,0)$, so, identifying the
contravariant and covariant indices through the Euclidean metric $g^{ij}$
(so that $M^{ij}=M^i_j=M_{ij}$), we obtain the following nonzero components of
the stress tensor:
$$
M^1_1=(E^1)^2+(B^1)^2-\frac12(\mathbf{E}^2+\mathbf{B}^2);
\ \ M^1_2=M^2_1=E^1\,E_2+B^1\,B_2;
$$
$$
M^2_2=(E^2)^2+(B^2)^2-\frac12(\mathbf{E}^2+\mathbf{B}^2); \ \
M^3_3=-\frac12(\mathbf{E}^2+\mathbf{B}^2).
$$
Since $M^1_1=-M^2_2$, the trace of $M$ is
$Tr(M)=-\frac12(\mathbf{E}^2+\mathbf{B}^2)$.

The eigen value equation acquires the simple
form
$$
\big[(M^1_1)^2-(\lambda)^2\big]+(M^1_2)^2\big](M^3_3-\lambda)=0.
$$
The corresponding eigen values are
$$
\lambda_1=-\frac12(\mathbf{E}^2+\mathbf{B}^2);\ \
\lambda_{2,3}=\pm\sqrt{(M^1_1)^2+(M^1_2)^2}= \pm\frac12\sqrt{(I_1)^2+(I_2)^2} ,
$$
where
$I_1=\mathbf{B}^2-\mathbf{E}^2,\, I_2=2\mathbf{E}.\mathbf{B}$.

The corresponding to
$\lambda_1$ eigen vector $Z_1$ must satisfy the equation
$$
\mathbf{E}(\mathbf{E}.Z_1)+\mathbf{B}(\mathbf{B}.Z_1)=0,
$$
 and since the non-zero
$(\mathbf{E},\mathbf{B})$ are lineary independent, the two coefficients
$(\mathbf{E}.Z_1)$ and $(\mathbf{B}.Z_1)$ must be equal to
zero, therefore, $Z_1\neq 0$ must be orthogonal to $\mathbf{E}$ and
$\mathbf{B}$, i.e. $Z_1$ must be colinear to $\mathbf{E}\times\mathbf{B}$:

The other two eigen vectors $Z_{2,3}$
satisfy correspondingly the equations
$$ \mathbf{E}(\mathbf{E}.Z_{2,3})+
\mathbf{B}(\mathbf{B}.Z_{2,3})=\Big[\pm\frac12\sqrt{(I_1)^2+(I_2)^2}+
\frac12(\mathbf{E}^2+\mathbf{B}^2)\Big]Z_{2,3}.         \ \ \ \ \ \    (*)
$$
Taking into account the easily verified relation
$$
\frac14\Big[(I_1)^2+(I_2)^2\Big]=
\left(\frac{\mathbf{E}^2+\mathbf{B}^2}{2}\right)^2-
|\mathbf{E}\times\mathbf{B}|^2 ,
$$
so that
$$
\frac{\mathbf{E}^2+\mathbf{B}^2}{2}- |\mathbf{E}\times\mathbf{B}|\geq 0 \ ,
$$
we conclude that the coefficient before $Z_{2,3}$ on the right is always different
from zero, therefore, the eigen vectors $Z_{2,3}(p)$ lie in the plane defined
by $(\mathbf{E}(p),\mathbf{B}(p)), \ p\in \mathbb{R}^3$. In particular,
the above mentioned transformation properties of the
Maxwell stress tensor $M(V,W)\rightarrow (a^2+b^2)M(V,W)$ show that the
corresponding eigen directions do not change under the transformation
$(V,W)\rightarrow (V\,a-W\,b,V\,b+W\,a)$.
\vskip 0.2cm
The above consideration suggests: {\it the intrinsically allowed
dynamical abilities of the field object might be: {\it translational}
along $(\mathbf{E}\times\mathbf{B})$, and {\it rotational} inside the plane
defined by $(\mathbf{E},\mathbf{B})$, hence, we may expect finding field
objects the propagation of which shows intrinsic local compatability between
rotation and translation.}
\vskip 0.2cm
It is natural to ask now {\bf under
what conditions the very $\mathbf{E}$ and $\mathbf{B}$ may be eigen vectors of
$M(\mathbf{E},\mathbf{B})$}? Assuming $\lambda_2=\frac12\sqrt{(I_1)^2+(I_2)^2}$
and $Z_2=\mathbf{E}$ in the above relation and having in view that
$\mathbf{E}\times\mathbf{B}\neq 0$ we obtain that
$\mathbf{E}(\mathbf{E}^2)+\mathbf{B}(\mathbf{E}.\mathbf{B})$ must be
proportional to $\mathbf{E}$, so, $\mathbf{E}.\mathbf{B}=0$,
i.e. $I_2=0$. Moreover, substituting now $I_2=0$ in that same  relation we obtain
$$
\mathbf{E}^2=\frac12(\mathbf{B}^2-\mathbf{E}^2)+
\frac12(\mathbf{E}^2+\mathbf{B}^2)=\mathbf{B}^2, \ \ \text{i.e.}, \ \ I_1=0.
$$
The case
"-" sign before the square root, i.e. $\lambda_3=-\frac12\sqrt{(I_1)^2+(I_2)^2}$,
leads to analogical conclusions just the role of $\mathbf{E}$ and $\mathbf{B}$
is exchanged.

\vskip 0.2cm
{\bf Corollary 1}. $\mathbf{E}$ and $\mathbf{B}$ may be eigen
vectors of $M(\mathbf{E},\mathbf{B})$ only if
 $I_1=I_2=0$.
\vskip 0.2cm

These considerations suggest that if $I_1=0$, i.e.
$|\mathbf{E}|^2=|\mathbf{B}|^2$, and propagation takes place, then the energy
density can be presented in terms of each of the two formal constituents,
moreover, in this respect, both constituents have the same rights. Therefore, a
local mutual energy exchange between any supposed two subsystems, formally
represented by appropriate combinations of $\mathbf{E},\mathbf{B}$, is not
forbidden in general, but, if it takes place, it must be {\it simultaneous} and
in {\it equal quantities}. Hence, if $I_1=0$ and
$I_2=2\mathbf{E}.\mathbf{B}=0$, internal energy redistribution between the two
supposed subsystems of the field object would be allowed, but such an exchange
should occur {\it without available interaction energy}.

The following question is also of interest: is it {\it physically} allowed to
interprit each of the two vector fields $\mathbf{E},\mathbf{B}$ {\it not as
formal constituents}, but as formal images of recognizable time-stable physical
{\it subsystems} of an electromagnetic field object?

Trying to answer this question we note that the relation
$\mathbf{E}^2+\mathbf{B}^2=2|\mathbf{E}\times\mathbf{B}|$ and the required
time-recognizability during propagation (with velocity "c") of each subsystem
of the field object suggest/imply also that {\it each of the two subsystems must
be able to carry locally momentum and to exchange locally momentum with the
other one}, since this relation means that the energy density is always
strongly proportional to the momentum density magnitude
$\frac1c|\mathbf{E}\times\mathbf{B}|$. Hence, the couple
$(\mathbf{E},\mathbf{B})$ is able to carry momentum, but {\it neither} of the
formal constituents $\mathbf{E},\mathbf{B}$ {\it is able to carry momentum
separately}. Moreover, the important observation here is that, verious
combinations constructed out of the formal constituents $\mathbf{E}$ and
$\mathbf{B}$, e.g., $(\mathbf{E}\,cos\theta-\mathbf{B}\,sin\theta,
\mathbf{E}\,sin\theta+\mathbf{B}\,cos\theta)$, where $\theta(x,y,z;t)$ is a
functon, may be considered as possible representatives of the two recognizable
subsystems since they carry the same energy
$\frac12(\mathbf{E}^2+\mathbf{B}^2)$ and momentum
$\frac1c|\mathbf{E}\times\mathbf{B}|$ densities.

We note also the following.

Let $(\mathbf{E},\mathbf{B})$ be nonzero and {\it lineary independent}, then
the triple $(\mathbf{E},\mathbf{B},\mathbf{E}\times\mathbf{B})$
defines a frame  and a  $g$-coframe at every point, where the field object is
different from zero.
We denote the corresponding frame by ${\cal R'}$, so we can write
${\cal R'}=(\mathbf{E},-\varepsilon\mathbf{B},
-\varepsilon\mathbf{E}\times\mathbf{B})$, where $\varepsilon=\pm 1$.

 Since the physical dimension of the third vector
$\mathbf{E}\times\mathbf{B}$ is different from that of the first two, we
introduce the factor $\gamma$ according to:
$$
\gamma=\frac{1}{\sqrt{\frac{\mathbf{E}^2+\mathbf{B}^2}{2}}}.
$$
Making use of $\gamma$, we introduce the so called {\it electromagnetic frame}
\index{electromagnetic frame}:
$$
{\cal R}=\left[\gamma\mathbf{E},-\varepsilon\gamma\mathbf{B},
-\varepsilon\gamma^2\mathbf{E}\times\mathbf{B}\right].
$$
Hence, at every point we've got two frames: ${\cal R}$, and the dimensionles
coordinate frame
${\cal R}_o=l_o\Bigl[{\partial_x},{\partial_y},{\partial_z}\Bigr]$,
$dim\,l_o=length$, as well as the corresponding co-frames ${\cal R}^*$ and
${\cal R}_o^*=l_o^{-1}(dx,dy,dz)$. The corresponding transformation matrix
${\cal M}:{\cal R}_o\rightarrow{\cal R}$ is defined by the relation ${\cal
R}_o.{\cal M}={\cal R}$.  So, we obtain $$ \mathcal{M}=\begin{vmatrix} \alpha
E^1&-\varepsilon\gamma B^1&-\varepsilon\gamma^2(\mathbf{E}\times\mathbf{B})^1\\
\gamma E^2&-\varepsilon\gamma
B^2&-\varepsilon\gamma^2(\mathbf{E}\times\mathbf{B})^2\\ \gamma
E^3&-\varepsilon\gamma B^3&-\varepsilon\gamma^2(\mathbf{E}\times\mathbf{B})^3
\end{vmatrix}. $$

Let's now see when the matrix $\mathcal{M}$ does not change the 3-volume
$\omega=dx\wedge dy\wedge dz$. Such a property requires
$det(\mathcal{M})=1$. Now, from linear algebra it is known that such matrices
have determinants equal to
$\gamma^4(\mathbf{E}\times\mathbf{B}).(\mathbf{E}\times\mathbf{B})=
[\gamma^2|\mathbf{E}||\mathbf{B}||sin(\mathbf{E},\mathbf{B})|]^2$.
So, this
requirenment reduces to $\gamma^2|\mathbf{E}\times\mathbf{B}|=1$. This last
equation is equivalent to
$$
|\mathbf{E}|^2-2|\mathbf{E}||\mathbf{B}||sin\beta|+|\mathbf{B}|^2=0,
$$
where $\beta$ is the angle between $\mathbf{E}$ and $\mathbf{B}$.
Expressing $|\mathbf{E}|$ as a function of $|\mathbf{B}|$ through solving
this quadratic equation with respect to $|\mathbf{E}|$, we obtain
$$
0<|\mathbf{E}|_{1,2}=|\mathbf{B}||sin\beta|\pm |\mathbf{B}|\sqrt{sin^2\beta-1}.
$$
This inequality is possible only if $|sin^2\beta|=1$, so,
$|\mathbf{E}|=|\mathbf{B}|,\ \mathbf{E}.\mathbf{B}=0$, and
$\sqrt{det(\mathcal{M})}=\gamma^2|\mathbf{E}||\mathbf{B}|$.

\noindent

{\bf Corollary 2.}
The {\it unimodular} nature of $\cal M$: $\det{\cal M}=1$,
requires $\mathbf{E}.\mathbf{B}=0, \mathbf{B}^2=\mathbf{E}^2$, so,
an orthonormal nature of the frame $\cal{R}$.

In our view these important properties have to be kept in mind when searching
 adequate equations, satisfied by the mathematical images of the
physical constituents of time dependent and space propagating electromagnetic
objects.

According to the obove considerations we may assume the following view on
{\it real}  electromagnetic field objects:
\vskip 0.3cm
{\bf Every real electromagnetic field object is built of two recognizable and
dynamically compatible subsystems $(A_1,A_2)$, the mathematical images of which
can be algebraically represented in terms of $(\mathbf{E},\mathbf{B})$, both
these subsystems carry always the same quantity of stress-energy-momentum,
guaranteeing in this way that, any mutual energy-momentum exchange between $A_1$
and $A_2$ will always be in equal quantities and simultanious}.

\section{Real electromagnetic field objects viewed as built of two
vector field constituents, being organized in two recognizable
and permanently interacting subsystems.}

\subsection{Some formal relations}

We begin with some notations and easily verified relations.
Let $g$ denote the euclidean metric on $\mathbb{R}^3$. We introduce the
following notations:
$$ \tilde{g}(\mathbf{E})=\eta, \ \
\tilde{g}(\mathbf{B})=\beta, \ \ \tilde{g}^{-1}(*\eta)=\bar{*\eta}, \ \
\tilde{g}^{-1}(*\beta)=\bar{*\beta}.
$$
Then we obtain
$$
\tilde{g}(\it{curl}\,\mathbf{E}\times\mathbf{E})=i(\mathbf{E})\mathbf{d}\eta,\
\ \tilde{g}(\it{curl}\,\mathbf{B}\times\mathbf{B})=i(\mathbf{B})\mathbf{d}\beta,
$$
$$
\tilde{g}(\it{curl}\,\mathbf{E}\times\mathbf{B})=i(\mathbf{B})\mathbf{d}\eta, \
\ \tilde{g}(\it{curl}\,\mathbf{B}\times\mathbf{E})=i(\mathbf{E})\mathbf{d}\beta
, $$ $$ \mathbf{E}\,\mathrm{div}(\mathbf{B})=i(\bar{*\eta})\mathbf{d}*\beta, \
\ \mathbf{B}\,\mathrm{div}(\mathbf{E})=i(\bar{*\beta})\mathbf{d}*\eta . $$
Further we shall use the notations and relations from Sec.2.

We introduce now some new relations.

Let  $E$ and $E^*$ be two dual real finite dimensional vector spaces.
The duality between $E$ and $E^*$ allows to distinguish the following
(anti)derivation. Let $h\in E$, then we obtain the derivation
$i(h)$, or $i_h$, in $\Lambda(E^*)$ of degree $(-1)$ (sometimes called
substitution/contraction/isertion operator, interior product, algebraic flow)
 according to:
$$
i(h)(x^{*1}\wedge\dots\wedge
x^{*p})=\sum_{i=1}^{p}(-1)^{(i-1)}\langle x^{*i},h\rangle
x^{*1}\wedge\dots\wedge\hat{x^{*i}}\wedge
\dots\wedge x^{*p}.
$$
Clearly, if $u^*\in \Lambda^p(E^*)$ and $v^*\in\Lambda(E^*)$ then
$$
i(h)(u^*\wedge v^*)=(i(h)u^*)\wedge v^*+(-1)^pu^*\wedge i(h)v^*.
$$
Also, we get
$$
i(h)u^*(x_1,\dots,x_{p-1})=u^*(h,x_1,\dots,x_{p-1}), \
$$
$$
i(x)\circ i(y)=-i(y)\circ i(x).
$$

This antiderivation is extended to a mapping $i(h_1\wedge\dots\wedge h_p):
\Lambda^m(E^*)\rightarrow\Lambda^{(m-p)}(E^*)$, $m\geqq p$,
according to
$$
i(h_1\wedge h_2\wedge\dots\wedge h_p)u^*=i(h_p)\circ\dots\circ i(h_1)\,u^*.
$$
Note that this extended mapping is not an antiderivation, except for $p=1$.

This mapping is extended to multivectors and exterior forms which are linear
combinations: if $\Psi=\Psi_1+\Psi_2+...$ is an arbitrary multivector on $E$ and
$\Phi=\Phi^1+\Phi^2+...$ is an arbitrary exterior form on $E^*$ then
$i_{\Psi}\Phi$ is defined as extention by linearity, e.g.,
$$
i(\Psi_1+\Psi_2)(\Phi^1+\Phi^2)
=i(\Psi_1)\Phi^1+i(\Psi_1)\Phi^2+i(\Psi_2)\Phi^1+i(\Psi_2)\Phi^2.
$$

This extension of the interior product allows to extend the Lie derivative of a
differential form $\alpha$ along a vector field $X$ to a
derivative of $\alpha$ along a multivector field $T$ [17], according to
$$
\mathcal{L}_T(\Phi)=
\mathbf{d}\circ i_T\Phi-
(-1)^{deg(T)}i_T\circ\mathbf{d}\Phi.
$$
If $\mathcal{L}_T(\Phi)=0$ this extension allows to consider $T$ as a symmetry
of $\alpha$.

 We construct now the $\varphi$-extended insertion operator. Let
$E_1$ and $E_2$ be two real vector spaces with corresponding bases ${e_i,
i=1,2,...,dimE_1}$ and ${k_j, j=1,2,...,dimE_2}$, $T=\mathfrak{t}^i\otimes e_i$
be a $E_1$-valued q-vector, $\Phi=\alpha^j\otimes k_j$ be a $E_2$-valued p-form
with $q\leq p$, and $\varphi:E_1\times E_2\rightarrow F$ be a bilinear map into
the vector space $F$. Now we define $i^{\varphi}_T\Phi\in\Lambda^{p-q}(M,F)$:
\begin{equation}
 i^{\varphi}_T\Phi= i_{\mathfrak{t}^i}\alpha^j\otimes\varphi(e_i,k_j), \ \ \
i=1,2,...,dim(E_1), \ j=1,2,...,dim(E_2).
\end{equation}
Also, if $T_1, T_2$ are two multivectors and $\alpha,\beta$ are two
forms then $(i\otimes i)_{T_1\otimes T_2}(\alpha\otimes \beta)$ is defined by
$$
(i\otimes i)_{T_1\otimes T_2}(\alpha\otimes \beta)=
i_{T_1}\alpha\otimes i_{T_2}\beta.
$$

We can define now the $\varphi$-extended Lie derivative. Let $M$ be a
$n$-dimensional manifold, $\Phi$ be a $E_1$-valued differential $p-$form on
$M$, $T$ be a $E_2$-valued $q$-multivectir field on $M$, with $q\leq p$ and
$\varphi:E_1\times E_2\rightarrow F$ be a bilinear map. The $\varphi$-extended
Lie derivative
$$
\mathcal{L}^{\varphi}_T:
\Lambda^p(M,E_1)\times\mathfrak{X}^q(M,E_2)\rightarrow\Lambda^{p-q+1}(M,F)
$$
is defined as follows [21]:
\begin{equation}
\mathcal{L}^{\varphi}_T(\Phi)=
\mathbf{d}\circ i^{\varphi}_T\Phi-
(-1)^{deg(T).deg(\mathbf{d})}i^{\varphi}_T\circ\mathbf{d}\Phi,
\end{equation}
where $\mathbf{d}$ is the exterior derivative on $M$, so, $deg(\mathbf{d})=1$.
This definition suggests to consider the tensor field $T$ as a {\it local}
$\varphi$-symmetry of the differential form $\Phi$ when
$\mathcal{L}^{\varphi}_T(\Phi)=0$.

\subsection{Static case}

We begin with the {\it strongly idealized} static case where the constituent
is modelled by a vector field on $\mathbb{R}^3$, denoted by
$\mathbf{E}$. In order to recognize this vector field among the other ones we
introduce 1-dimensional vector space $V_o$, its dual $V_o^*$, with corresponding
dual bases ${e}$ and ${\varepsilon}$, so our formal representation of the
constituent looks as $\mathbf{E}\otimes e$. In searching for a partner
our field $\mathbf{E}$ defines its
$g$-dual 1-form $\eta=\tilde{g}(\mathbf{E})$, and making use of
the Hodge star $*$ defined by $g$, it  finds its partner constituent in
 the 2-form $*\eta$, which is equal to $i_{\mathbf{E}}\omega$,
$\omega=dx\wedge dy\wedge dz$. According to the above view the balance in this
idealized case between $\mathbf{E}\otimes e$ and $*\eta\otimes\varepsilon$
should be given by
$$
\mathcal{L}^\varphi_{(\mathbf{E}\otimes e)}(\eta\otimes\varepsilon)=
\mathcal{L}^\varphi_{(\bar{*\eta}\otimes e)}(*\eta\otimes\varepsilon),
$$
where $\varphi$ in this case is just the coupling
$\langle\varepsilon,e\rangle=1$ between $e$ and $\varepsilon$. Expanding this
relation we get
$$
\mathbf{d}\langle\eta,\mathbf{E}\rangle+i_{\mathbf{E}}\mathbf{d}\eta=
\mathbf{d}\langle*\eta,\bar{*\eta}\rangle-i_{\bar{*\eta}}\mathbf{d}*\eta.
$$
Since $\langle\eta,\mathbf{E}\rangle=\langle*\eta,\bar{*\eta}\rangle$ we get
$$
i_{\mathbf{E}}\mathbf{d}\eta=-i_{\bar{*\eta}}\mathbf{d}*\eta,\ \ \ \  i.e., \ \
\ \ \ \ i_{\mathbf{E}}\mathbf{d}\eta+i_{\bar{*\eta}}\mathbf{d}*\eta=0,
$$
which is, according to relation (3), just the zero value of the divergence
of the determined by $\mathbf{E}$ Maxwell stress tensor: The flow of
$\mathbf{E}$ across $\mathbf{d}\eta$ is balanced by the flow of $\bar{*\eta}$
across $\mathbf{d}*\eta$.

This very elementary example suggests that, even {\it only one} vector field,
$\mathbf{E}$ in our case, in order to {\it survive as a static stress generating
factor} in the 3-space, looks for a balancing
parthner, it finds such one in $\bar{i_{\mathbf{E}}}\omega=\bar{*\eta}$, its
 $\tilde{g}$-images, and the {\it static} stress flow
$i_{\bar{*\eta}}\mathbf{d}*\eta$.

 \vskip 0.3cm We pass now to the
case of {\it two} static stress generating vector fields, denoted by
$\mathbf{E}$ and $\mathbf{B}$. The new moment now is that the resulted
generated local stress, although {\it static}, may depend on the {\bf mutual}
influence between the generated two local stresses by each one of the fields.
\vskip 0.2cm
According to the above notations the two vector fields
$\mathbf{E}$ and $\mathbf{B}$ appear together with their co-vectors $\eta$ and
$\beta$. Now the 1-dimensional vector space $V_o$ should be, naturally,
replaced by a 2-dimensional vector space $V$ and its dual $V^*$, euclidean
metric $h$, and corresponding $h$-dual bases $\{e_1,e_2\}$ and
$\{\varepsilon_1,\varepsilon_2\}$: $\langle
\varepsilon^i,e_j\rangle=\delta^i_j, \ i,j=1,2$.  So, $(\mathbf{E},\mathbf{B})$
define a subsystem $\bar{\Omega}$ by
$$
\bar{\Omega}=\mathbf{E}\otimes e_1+\mathbf{B}\otimes e_2 ,\ \  \text{ and its
$g$-dual co-image}\ \ \ \Omega=\eta\otimes e_1+\beta\otimes e_2.
$$
Since now the volume form in $V^*$ is essential and is given by
$\varepsilon^1\wedge\varepsilon^2$, we are going to introduce the balancing
partner field $\Sigma$ in two steps. First, the flow of $\bar{\Omega}$ across
$\omega\otimes\varepsilon^1\wedge\varepsilon^2$:
$$
 \Sigma'=i(\bar{\Omega})(\omega\otimes\varepsilon^1\wedge\varepsilon^2).
$$
Explicitly
$$
\Sigma'=i(\mathbf{E}\otimes e_1+\mathbf{B}\otimes e_2)
(\omega\otimes\varepsilon^1\wedge\varepsilon^2)=
i(\mathbf{E})\omega\otimes i(e^1)(\varepsilon^1\wedge\varepsilon^2)+
i(\mathbf{B})\omega\otimes i(e^2)(\varepsilon^1\wedge\varepsilon^2)
$$
$$
=i(\mathbf{E})\omega\otimes\varepsilon^2-i(\mathbf{B})\omega\otimes\varepsilon^1
=-*\beta\otimes\varepsilon^1+*\eta\otimes\varepsilon^2.
$$
Now $\Sigma$ is defined by passing to $V$-valued 2-form by
$$
\Sigma=(id\otimes\tilde{h}^{-1})(\Sigma')=-*\beta\otimes e_1+*\eta\otimes e_2.
$$
Finally, the balancing partner is represented by $\Sigma$ and
$$
\bar{\Sigma}=-\bar{*\beta}\otimes e_1+\bar{*\eta}\otimes e_2.
$$

Now the corresponding local {\it static} stress balance relation must pay due
respect to the way the two stress generating formal constituents $\mathbf{E}$ and
$\mathbf{B}$ generate interaction: the interaction must
 take care of their identities through recognizing them as eigen
vectors of the stress-energy tensor, so, they should carry the {\it same
local stress}, therefore, their {\it static} exchange stress, i.e., stress
balance, must be {\it simultaneous} and {\it in equal quantities} (Corollary 1).
In view of this, paying due respect to these properties of mutual symmetry and
compatibility, we choose $\varphi$ to be the {\it symmetrized tensor
product} denoted by $"\vee"$, and write:
 \begin{equation}
\mathcal{L}^{\vee}_{\bar{\Omega}}(\Omega)=
\mathcal{L}^{\vee}_{\bar{\Sigma}}(\Sigma).
\end{equation}
We obtain:
$$
\mathcal{L}^{\vee}_{\bar{\Omega}}(\Omega)=
[\mathbf{d}\langle\eta,\mathbf{E}\rangle+
i(\mathbf{E})\mathbf{d}\eta]\otimes e_1\vee e_1+
[\mathbf{d}\langle\beta,\mathbf{B}\rangle+
i(\mathbf{B})\mathbf{d}\beta]\otimes e_2\vee e_2
$$
$$
+[\mathbf{d}\langle\eta,\mathbf{B}\rangle+
i(\mathbf{B})\mathbf{d}\eta+\mathbf{d}\langle\beta,\mathbf{E}\rangle+
i(\mathbf{E})\mathbf{d}\beta]\otimes e_1\vee e_2.
$$
\vskip 0.3cm
$$
\mathcal{L}^{\vee}_{\bar{\Sigma}}(\Sigma)=
[\mathbf{d}\langle *\beta,\bar{*\beta}\rangle-
i(\bar{*\beta})\mathbf{d}*\beta]\otimes e_1\vee e_1+
[\mathbf{d}\langle *\beta,\bar{*\beta}\rangle-
i(\bar{*\eta})\mathbf{d}*\eta]\otimes e_2\vee e_2
$$
$$
+[-\mathbf{d}\langle *\eta,\bar{*\beta}\rangle+
i(\bar{*\beta})\mathbf{d}*\eta-
\mathbf{d}\langle *\beta,\bar{*\eta}\rangle+
i(\bar{*\eta})\mathbf{d}*\beta]\otimes e_1\vee e_2 .
$$

So, the balance relation (7) gives the following three equations:
$$
\mathbf{d}\langle\eta,\mathbf{E}\rangle+i(\mathbf{E})\mathbf{d}\eta=
\mathbf{d}\langle *\beta,\bar{*\beta}\rangle-i(\bar{*\beta})\mathbf{d}*\beta,
$$
$$
\mathbf{d}\langle\beta,\mathbf{B}\rangle+i(\mathbf{B})\mathbf{d}\beta=
\mathbf{d}\langle *\eta,\bar{*\eta}\rangle-i(\bar{*\eta})\mathbf{d}*\eta,
$$
$$
\mathbf{d}\langle\beta,\mathbf{E}\rangle+i(\mathbf{E})\mathbf{d}\beta+
\mathbf{d}\langle\eta,\mathbf{B}\rangle+i(\mathbf{B})\mathbf{d}\eta=
-\mathbf{d}\langle *\eta,\bar{*\beta}\rangle-
\mathbf{d}\langle *\beta,\bar{*\eta}\rangle+i(\bar{*\beta})\mathbf{d}*\eta+
i(\bar{*\eta})\mathbf{d}*\beta.
$$
In view of the relations
$$
\langle\eta,\mathbf{E}\rangle=g(\mathbf{E},\mathbf{E})=\mathbf{E}^2, \
\langle\beta,\mathbf{B}\rangle=g(\mathbf{B},\mathbf{B})=\mathbf{B}^2, \
\langle *\beta,\bar{*\beta}\rangle=\mathbf{B}^2, \
\langle *\eta,\bar{*\eta}\rangle=\mathbf{E}^2, \
$$
$$
\langle\eta,\mathbf{B}\rangle=g(\mathbf{E},\mathbf{B})=\mathbf{E}.\mathbf{B}, \
\langle\beta,\mathbf{E}\rangle=g(\mathbf{B},\mathbf{E})=\mathbf{B}.\mathbf{E}, \
\langle *\beta,\bar{*\eta}\rangle=\mathbf{B}.\mathbf{E}, \
\langle *\eta,\bar{*\beta}\rangle=\mathbf{E}.\mathbf{B},
$$
the equations read
$$
i(\mathbf{E})\mathbf{d}\eta+i(\bar{*\beta})\mathbf{d}*\beta
=\mathbf{d}\langle\mathbf{B}^2-\mathbf{E}^2\rangle
$$
$$
i(\mathbf{B})\mathbf{d}\beta+i(\bar{*\eta})\mathbf{d}*\eta
=\mathbf{d}\langle\mathbf{E}^2-\mathbf{B}^2\rangle,
$$
$$
i(\mathbf{E})\mathbf{d}\beta+i(\mathbf{B})\mathbf{d}\eta-
i(\bar{*\eta})\mathbf{d}*\beta-i(\bar{*\beta})\mathbf{d}*\eta
=-\mathbf{d}(\mathbf{4\,E.B})
$$
From the first two equations it follows the equation, i.e., {\it the static
balance equation},
$$
i(\mathbf{E})\mathbf{d}\eta+i(\bar{*\beta})\mathbf{d}*\beta+
i(\mathbf{B})\mathbf{d}\beta+i(\bar{*\eta})\mathbf{d}*\eta=0,
$$
which is the Maxwell local conservation law
$\nabla_iM^i_j(\mathbf{E},\mathbf{B})=0$ for the stress tensor
$$
M^i_j(\mathbf{E},\mathbf{B})=
 \mathbf{E}^i\mathbf{E}_j+\mathbf{B}^i\mathbf{B}_j-
\frac12(\mathbf{E}^2+\mathbf{B}^2)\delta^i_j
$$
$$
=\frac12\Big
[\mathbf{E}^i\mathbf{E}_j+\big(\tilde{g}^{-1}\circ*\tilde{g}(\mathbf{E})\big)^{ik}
\big(*\tilde{g}(\mathbf{E})\big)_{kj}+
\mathbf{B}^i\mathbf{B}_j+\big(\tilde{g}^{-1}\circ*\tilde{g}(\mathbf{B})\big)^{ik}
\big(*\tilde{g}(\mathbf{B})\big)_{kj}\Big ] .
$$
Recalling (Sec.3) that if we require the vector fields $\mathbf{E}$ and
$\mathbf{B}$ to define at every point {\it eigen directions} of
$M(\mathbf{E},\mathbf{B})$ and {\it unimodular} nature of the generated
electromagnetic matrix $\mathcal{M}$,
 $\mathcal{M}^*\omega=\omega$, where $\omega=dx\wedge dy\wedge dz$,
then we should assume
$$
\mathbf{E}^2=\mathbf{B}^2 \ \ \text{and} \ \ \ \mathbf{E.B}=0.
$$
It is now elementary to see that under these last assumptions our {\it static}
balance relation $\mathcal{L}^{\vee}_{\bar{\Omega}}(\Omega)=
\mathcal{L}^{\vee}_{\bar{\Sigma}}(\Sigma)$ reduces to
\begin{equation}
i^{\vee}_{\bar{\Omega}}\mathbf{d}\Omega=-
i^{\vee}_{\bar{\Sigma}}\mathbf{d}\Sigma \,.
\end{equation}
This suggests to consider $\bar{\Omega}, \bar{\Sigma}$, or $\Omega,\Sigma$, as
formal images of two {\it subsystems} of the object considered, which subsystems
demonstrate stable stress equilibrium: {\it any stress lost by the first one is
fully accepted by the second one and vice versa}. This corresponds to the fact
that there is NO interaction stress in $M^i_j(\mathbf{E},\mathbf{B})$: the
whole stress is sum of the stresses carryied by $\Omega$ and $\Sigma$. The
hidden "dynamical" aspect of this static equilibrium is clearly seen from the
reduced three equations.
$$
i(\mathbf{E})\mathbf{d}\eta+i(\bar{*\beta})\mathbf{d}*\beta=0 ,
$$
$$
i(\mathbf{B})\mathbf{d}\beta+i(\bar{*\eta})\mathbf{d}*\eta=0,
$$
$$
i(\mathbf{E})\mathbf{d}\beta+i(\mathbf{B})\mathbf{d}\eta-
i(\bar{*\eta})\mathbf{d}*\beta-i(\bar{*\beta})\mathbf{d}*\eta=0 .
$$

 \vskip 0.3cm

\subsection{Time dependent case}

First we note that introducing {\it time} is considered here as a quantitative
comparing the courses of two physically independent processes, the one of which
we call {\it referent}, e.g., the progress of appropriate watch, then the other
one attains significance of {\it parametrised} process.

Hence, we have to specially note that the {\it time} parameter $t$ used in this
subsection we consider as {\it external} to the spatial coordinates $(x,y,z)$
parameter, and the corresponding referent process must NOT influence the
parametrised process . Some formal consequences of this consideration should be
noted:

- time-derivatives are NOT derivatives along $\mathbb{R}^3$-spatial vector fields
and, by assumption, corresponding local commutation relation between
$\frac{\partial}{\partial t}$ and
$\big{(}\frac{\partial}{\partial x}, \frac{\partial}{\partial y},
\frac{\partial}{\partial z}\big{)}$
{\it always} holds,

- time-derivatives do {\it not} change the tensor nature of the differentiated
object.

Naturally, from physical viewpoint, any observed time change of
the above discussed stress balance in the {\it static} case should presume
corresponding {\it influence}, leading to its
violation, and, of course, of its formal representation - relation (8).
Physically, it may be expected the electromagnetic field object described, to
sirvive through some kind of {\it time "pulsating"} at the space points, or
through a {\it propagation as a whole} in the 3-space, or, both. So: {\it the
local static balance should be replaced by an appropriate intrincally
compatible local dynamical and time dependent balance}. Hence, in order to
survive, our object must be able to generate {\it appropriate} spatial changes
inside any occupied spatial area. In particular, in order the eigen nature of
the static stress tensor to be approprately kept as eigen nature of the new
"propagational" stress-energy-momentum tensor, the two zero divergences
$*\mathbf{d}*\eta=0, *\mathbf{d}*\beta=0$, might be not necessarily kept to
hold, as the explicit form of static equations (8) at the end of the previous
subsection allow.

To this time-dependence of the behaviour of our electromagnetic field oject we
are going to give formal decription by means of finding appropriate change of
the static equation (8).

Equation (8) formally postulates equivalence between two vector valued 1-forms,
so, any introduced influence object, representing how the  new time-dependent
balance would look like, is expected, formally, also to be 1-form, containing
appropriately first order $t$-derivative(s) and valued in the same vector space.
This allows a natural return to the static balance equation through setting
this new oject equal to zero.

Also, since the available spatial differential operators in (8) are just of {\it
first order}, it seems natural the corresponding formal influence object to
contain time derivatives of {\bf not higher} than first order. Clearly, in view
of the flow nature of the objects across their own spatial change objects in
the static relation (8), the influence object is expected to
express formally also a flow, but a flow across {\it time
differentiated} object. Moreover, it should be expected also this time
dependence to generate direct mutual influence between the two now
time-dependent subsystems. Finally, since time derivation must not change the
tensor nature of the differentiated object, and since $\Omega$ is 1-form, then
the 2-form $\Sigma$ is the natural candidate to be {t}-differentiated. So, we
may write
\begin{equation} i^{\vee}_{\bar{\Omega}}\mathbf{d}\Omega+
i^{\vee}_{\bar{\Sigma}}\mathbf{d}\Sigma =
\frac{1}{c}\,i^\vee_{\bar{\Omega}}\frac{\partial}{\partial t}\Sigma .
\end{equation}
Denoting $ct=\xi$, this equation (9) gives the following three equations
$$
i(\mathbf{E})\mathbf{d}\eta+i(\bar{*\beta})\mathbf{d}*\beta=
-i(\mathbf{E})\left(\frac{\partial}{\partial \xi}*\beta\right) ,
$$
$$
i(\mathbf{B})\mathbf{d}\beta+i(\bar{*\eta})\mathbf{d}*\eta=
i(\mathbf{B})\left(\frac{\partial}{\partial \xi}*\eta\right),
$$
$$
i(\mathbf{E})\mathbf{d}\beta+i(\mathbf{B})\mathbf{d}\eta-
i(\bar{*\eta})\mathbf{d}*\beta-i(\bar{*\beta})\mathbf{d}*\eta=
 i(\mathbf{E})\left(\frac{\partial}{\partial \xi}*\eta\right)-
i(\mathbf{B})\left(\frac{\partial}{\partial \xi}*\beta\right).
$$
\vskip 0.3cm

Having in view the expressions for the extended Lie derivatives  we can rewrite
these equatios as follows:
$$
L_{\mathbf{E}}\eta-\mathbf{d}\langle\eta,\mathbf{E}\rangle-
\big[L_{\bar{*\beta}}*\beta-\mathbf{d}\langle*\beta,\bar{*\beta}\rangle\big{]}=
-i(\mathbf{E})\left(*\frac{\partial\beta}{\partial\xi}\right)
$$
$$
L_{\mathbf{B}}\beta-\mathbf{d}\langle\beta,\mathbf{B}\rangle-
\big[L_{\bar{*\eta}}*\eta-\mathbf{d}\langle*\eta,\bar{*\eta}\rangle\big]=
i(\mathbf{B})\left(*\frac{\partial\eta}{\partial\xi}\right)
$$
$$
L_{\mathbf{E}}\beta-\mathbf{d}\langle\beta,\mathbf{E}\rangle+
L_{\mathbf{B}}\eta-\mathbf{d}\langle\eta,\mathbf{B}\rangle+
L_{\bar{*\eta}}*\beta-\mathbf{d}\langle*\beta,\bar{*\eta}\rangle+
L_{\bar{*\beta}}*\eta-\mathbf{d}\langle*\eta,\bar{*\beta}\rangle
$$
$$
=i(\mathbf{E})\left(*\frac{\partial\eta}{\partial\xi}\right)
-i(\mathbf{B})\left(*\frac{\partial\beta}{\partial\xi}\right).
$$

\subsection{Space-time representation}
In the frame of the space-time view on physical processes the introduced
variable $\xi=ct$ is no more indepentent on the choice of physical frames with
respect to which we introduce spatial coordinates and write down time-dependent
formal relations. Now $\xi$ is considered as appropriate coordinate, it
generates local coordinate base vetor $\frac{\partial}{\partial \xi}$ and
corresponding co-vector (or 1-form)
$d\xi, \langle d\xi,\frac{\partial}{\partial \xi}\rangle=1$.
So, the 3-volume $\omega=dx\wedge
dy\wedge dz$ naturally becomes a 3-form on the 4-dimensional spase-time
$\mathbb{R}^4$, and is extended to the 4-volume $\omega_o=dx\wedge dy\wedge
dz\wedge d\xi$. Our purpose now is to find appropriate 4-dimensional form of
our balance law given by equation (9).

  Recall our two basic objects: the vector valued differential 1-form
$\Omega=\eta\otimes e_1+\beta\otimes e_2$ and the vector valued differential
2-form $\Sigma=-*\beta\otimes e_1+*\eta\otimes e_2$ being defined entirely in
terms of objects previously introduced on $\mathbb{R}^3$. We want now
these objects to depend on $\xi$ as they depend on the spatial coordinates,
 so to be appropriately extended to objects on $\mathbb{R}^4$.

Note that the 2-form $\Sigma$ is defined making use of the 1-form $\Omega$ and
the 3-form $\omega=dx\wedge dy\wedge dz$. Now, the 4th dimension $\xi$
generates the coordinate 1-form $d\xi$, so,
$\Omega$ turns to $d\xi$ for help to extend to a 2-form on $\mathbb{R}^4$,
which is done in the simplest way: $\Omega\rightarrow\Omega\wedge d\xi$. We are
in position now to consider the difference $\Omega\wedge d\xi-\Sigma$. $$
\Omega\wedge
d\xi-\Sigma=(\eta\wedge d\xi)\otimes e_1+(\beta\wedge d\xi)\otimes
e_2+*\beta\otimes e_1-*\eta\otimes e_2
$$
$$
=(*\beta+\eta\wedge d\xi)\otimes e_1-(*\eta-\beta\wedge d\xi)\otimes e_2.
$$
In this way we get two differential 2-forms on $\mathbb{R}^4$ naturally
recognized by the basis vectors of the external vector spase $V$:
 $$
F=*\beta+\eta\wedge d\xi, \ \ \text{and} \ \ \
-G=*\eta-\beta\wedge d\xi ,
 $$
moreover, these two 2-forms are clearly identified as vector components of {\it
one} $V$-valued 2-form:
$$
 \mathbf{\Omega}=F\otimes e_1+G\otimes e_2.
$$

In order to define corresponding flow, as we did it in previous subsections,
we have to construct $\bar{\mathbf{\Omega}}$. The corresponding 2-vectors
$\bar{F}$ and $\bar{G}$ are easily introduced making use of the isomorphism
between 2-forms and 2-vectors defined by the volume form
$\omega_o=dx\wedge dy\wedge dz \wedge d\xi$ according to
$$
G=-i(\bar{F})\omega_o, \ \ \ F=i(\bar{G})\omega_o: \ \ \rightarrow \ \ \
\bar{\mathbf{\Omega}}=\bar{F}\otimes e_1+\bar{G}\otimes e_2.
$$

Another approach is to try to find appropriate {\it linear} map
$\psi:\Lambda^2(\mathbb{R}^4)\rightarrow \Lambda^2(\mathbb{R}^4)$ sending $F$
to $G$. So, we write down the presumed linear equation
$\psi(F)=G$:
$$ \psi(i_{\mathbf{B}}\omega + \eta\wedge d\xi)=
i_{\mathbf{E}}\omega-\beta\wedge d\xi.
$$
The linear nature of this presumed equation allows to
reduce now $\psi$ to the basis vectors of $\Lambda^2(\mathbb{R}^4):
(dx\wedge dy, dx\wedge dz, dy\wedge dz, dx\wedge d\xi, dy\wedge d\xi,
dz\wedge d\xi)$, which gives:
\[ \begin{array}{ll}
\psi(dx\wedge dy)=-dz\wedge d\xi &\psi(dx\wedge d\xi)=dy\wedge dz \\
\psi(dx\wedge dz)=dy\wedge d\xi &\psi(dy\wedge d\xi)=-dx\wedge dz \\
\psi(dy\wedge dz)=-dx\wedge d\xi &\psi(dz\wedge d\xi)=dx\wedge dy.
\end{array} \]
Obviously, $\psi$ must satisfy the condition
$\psi\,\circ \,\psi=-id_{\Lambda^2(\mathbb{R}^4)}$. Clearly, such linear map
should define {\it complex structure} in the space
$\Lambda^2(\mathbb{R}^4)$.  As is well known, the Hodge star operator $*$ in
Minkowski space-time, is defined by the relation
$\alpha\wedge *\beta=(-1)^{ind(\mathbf{g})}\mathbf{g} (\alpha,\beta)\omega_o$,
where $\omega_o=dx\wedge dy\wedge dz\wedge d\xi$,
 $\alpha,\beta$ are forms of the same rank, $ind(\mathbf{g})$ specifies
the number of minuses in canonical coordinates of the pseudometric used.
In our case the
Minkowski pseudometric $\mathbf{g}$ has in canonical coordinates the
components:  $\mathbf{g}_{\mu\mu}=(-1,-1,-1,1); \ \mathbf{g}_{\mu\nu}=0,
\mu\neq\nu=1,2,3,4$. It should be noted here, that an interior product
$i_X\omega_o$ is not always equal to $*_{\mathbf{g}}\mathbf{g}(X)$, where $X$
is a multivector. In our case of Minkowski space-time with this pseudo-metric
it is easy to verify that for 2-forms and $\mathbf{g}$-corresponding
2-vectors we obtain:
\begin{equation}
*F=-i(\bar{F})\omega_o , \ \ \  F=i(\bar{*F})\omega_o , \ \ \
\text{i.e.}, \ \ \  \bar{G}=\bar{*F}.
\end{equation}
In view of this, further we may use any of these two expressions.

 We turn now to the corresponding balance law. In view of the preliminary
assumed relations $\mathbf{E}^2=\mathbf{B}^2,
\mathbf{E}.\mathbf{B}=0$, i.e., $F\wedge *F=F\wedge F=0$, it reeds
\begin{equation}
i^{\vee}_{\bar{\mathbf{\Omega}}}\mathbf{d}\mathbf{\Omega}=0, \ \ \
\bar{\mathbf{\Omega}}=(\tilde{\mathbf{g}})^{-1}(\mathbf{\Omega}),
\end{equation}
i.e., the $\vee$-flow of ${\bar{\mathbf{\Omega}}}$ across its change
$\mathbf{d}\mathbf{\Omega}$ does NOT lead to losses. It has to be noted,
however, that this balance law may be written down without making use of
(pseudo)metric, the volume form $\omega_o$ serves sufficiently well.

Since now $G=*F=-i(\tilde{\mathbf{g}}^{-1}F)\omega_o=-i(\bar{F})\omega_o$,
equation (11) gives the following three equations
\begin{equation}
 i_{\bar{F}}\mathbf{d}F=0, \ \ \
i_{\bar{*F}}\mathbf{d}*F=0, \ \ \
i_{\bar{F}}\mathbf{d}*F+i_{\bar{*F}}\mathbf{d}F=0.
\end{equation}

 If $\delta=*\mathbf{d}*$ is the corresponding coderivative operator on
Minkowski spacetime, the first two equations of (12) are correspondingly
equivalent to
$$
(*F)_{\mu\nu}(\delta *F)^{\nu}=0, \ \ F_{\mu\nu}(\delta
F)^{\nu}=0,
$$
so, all nonlinear solutions: $\delta F\neq 0, \delta *F\neq
0$ of these equations must satisfy $det||F_{\mu\nu}||=0$, i.e.,
$\mathbf{E}.\mathbf{B}=0$, and together with the third equation

$$
i_{\bar{F}}\mathbf{d}*F+i_{\bar{*F}}\mathbf{d}F=
i(\bar{\delta *F})F+i(\bar{\delta F})(*F)=0
$$
this requirement for nonlinearity extends to $det||F\pm *F||=0$, which is
equivalent to $\mathbf{E}^2=\mathbf{B}^2$.

The first two
equations appeared first in [18], and the third jooined later in [19,20].

It is easy now to varify  that writing down (9) and (12) totally
in terms of $(\mathbf{E},\mathbf{B})$ we shall obtain the equations given at
the end of the previous subsection.

The corresponding stress-energy-momentum tensor of any solution of (12)
$$
T_{\mu}\,^{\nu}=-\frac12\big[F_{\mu\sigma}F^{\nu\sigma}+
(*F)_{\mu\sigma}(*F)^{\nu\sigma}\big]
$$
clearly notifies absence of interaction stress-energy between the two subsystems
formally represented by $F$ and $*F$: the whole stress-energy is the sum of
these quantities carried by $F$ and $*F$. Of course, this admits local exchange
$F\leftrightarrow *F$
of these quantities of special kind: {\it simultanious} and {\it in equal
quantities}.

 Here is a special class of nonlinear solutions of (12):
\begin{eqnarray*}
&&F=\varepsilon
u\,dx\wedge dz + u\,dx\wedge d\xi + \varepsilon p\,dy\wedge dz + p\,dy\wedge
d\xi \\
&&*F=-p\,dx\wedge dz - \varepsilon p\,dx\wedge d\xi + u\,dy\wedge dz +
\varepsilon u\,dy\wedge d\xi,
\end{eqnarray*} where
$$
u=\Phi(x,y,\xi+\varepsilon z)\cos\left(-\varepsilon
\kappa\frac{z}{\mathcal{L}_o}+const\right),
$$
$$ p=\Phi(x,y,\xi+\varepsilon
z)\sin\left(-\varepsilon \kappa\frac{z}{\mathcal{L}_o}+const\right), \ \
\mathcal{L}_o=const, \ \ \ \varepsilon=\pm 1, \ \ \ \kappa=\pm 1,
$$
and $\Phi$ is {\it arbitrary} function of it's four arguments.
The energy density of these solutions is given by $\Phi^2$, so, spatially
finite solutions are {\it allowed}. These solutions
 propagate along the coordinate $z$ inside some spatially infinite {\it helical
cylinder} and "rotate" left or right depending on the sign of the constant
$\kappa$. Their length size along the direction of propagation is
$2\pi\mathcal{L}_o$.

\vskip 0.5cm

{\bf Remarks and Comments}
\vskip 0.3cm
Recalling relation (6) and the zero values of
$$
F\wedge F=i_{\bar{F}}(*F)\omega_o=2\mathbf{E.B}\omega_o=0,
\ \ \ \ \text{and} \ \ \ \
F\wedge *F=-i_{\bar{F}}F\omega_o=(\mathbf{E}^2-\mathbf{B}^2)\omega_o=0,
$$
we see that the first two equations of (12)
clearly suggest to consider $\bar{F}$ es a
local symmetry of the 2-form $F$, but NOT as a local symmetry of $*F$, as well
as to consider $\bar{*F}$ as a local symmetry of the 2-form $*F$ but NOT as a
local symmetry of $F$:
$$
\mathcal{L}_{\bar{F}}F=\mathbf{d}(i_{\bar{F}}F)-
(-1)^{2}i_{\bar{F}}\mathbf{d}F=-i_{\bar{F}}\mathbf{d}F =0,
$$
$$
\mathcal{L}_{\bar{*F}}*F=\mathbf{d}(i_{\bar{*F}}*F)-
(-1)^{2}i_{\bar{*F}}\mathbf{d}*F=-i_{\bar{*F}}\mathbf{d}*F=0 .
$$
Also, since $\mathbf{d}\omega_o=0$,  we obtain
$$
\mathcal{L}_{\bar{F}}\omega_o=
\mathbf{d}(i_{\bar{F}}\omega_o)-(-1)^{2}i_{\bar{F}}\mathbf{d}\omega_o
=-\mathbf{d}*F,
\ \ \ \
\mathcal{L}_{\bar{*F}}\omega_o=
\mathbf{d}(i_{\bar{*F}}\omega_o)-(-1)^{2}i_{\bar{*F}}\mathbf{d}\omega_o
=\mathbf{d}F .
$$
The last relations give some other view on the relativistic form of
Maxwell free field equations
$\mathbf{d}F=0, \mathbf{d}*F=0$ : the two
{\it null} bivectors $\bar{F}$ and $\bar{*F}$ are local symmetries of the
standard volume form $\omega_o$ on Minkowski spacetime, in this sense, these
two relativistic Maxwell equations appear as extensions of the nonrelativistic
equations $div\mathbf{E}=0, div\mathbf{B}=0$, the invariant sense of which is
$$
L_{\bar{\mathbf{E}}}(dx\wedge dy\wedge dz)=0, \ \ \ \
 L_{\bar{\mathbf{B}}}(dx\wedge
dy\wedge dz)=0,
$$ i.e., the two fields $(\mathbf{E},\mathbf{B})$ do not change
locally the 3-volume.  Moreover, if we follow modern {\it guage} formulation of
relativistic charge-free Maxwell equations, then $\mathbf{d}F=0$ is in advance
assumed, and in view of relations (10) the only aditional equation should read
$\mathcal{L}_{\bar{F}}\omega_o=\mathbf{d}(i_{\bar{F}}\omega_o)=
-\mathbf{d}*F=0$.

Turning back to the static {\it null field} case where Maxwell free field
equations require $\it{curl}\,\mathbf{E}=\it{curl}\,\mathbf{B}=0$, and recalling
relations in subsec.4.1, we see that  the first two nonlinear static equations
at the end of subsec.4.2 allow $\it{curl}\,\mathbf{E}\neq 0, \ \it{curl}\,\mathbf{B}\neq
0$, i.e., $\mathbf{d}\eta\neq 0, \ \mathbf{d}\beta\neq 0$,
 since the 3d-matrices  $\mathbf{d}\eta$ and $\mathbf{d}\beta$ are
antisymmetric, and their determinants are {\it necessarily equal to zero}, which
allows the components of $\mathbf{E}$ and $\mathbf{B}$ to be
algebraically determined as nonzero functions of their derivatives, although
$div\mathbf{E}=div\mathbf{B}=0$ .

This allows in principle to consider the two Frobenius integrability conditions
$$
\mathbf{d}\eta\wedge\eta=(\mathbf{E}.\it{curl}\,\mathbf{E})\,\omega=0,
\ \ \ \ \mathbf{d}\beta\wedge\beta=(\mathbf{B}.\it{curl}\,\mathbf{B})\,\omega=0
$$
as compatible
with the nonlinear static equations, so, static electric and magnetic {\it
helicities}, which are not allowed by Maxwell static equations, not to be
excluded from the very beginning. For example, the vector fields $X$,
satisfuing $X\times\it{curl}\,X =0, div\,X=0$, known as Beltrami vector fields,
exist and are of definite interest in fluid mechanics and optics [22].

Following this line of consideration we find
$$
L_{\mathbf{E}\times\mathbf{B}}\,\omega=\mathbf{d}i_{\mathbf{E}\times\mathbf{B}}\,\omega+
i_{\mathbf{E}\times\mathbf{B}}\mathbf{d}\,\omega=
\mathbf{d}i_{\mathbf{E}\times\mathbf{B}}\,\omega=
div({\mathbf{E}\times\mathbf{B}})\,\omega=
(\mathbf{B}.\it{curl}\,\mathbf{E}-\mathbf{E}.\it{curl}\,\mathbf{B})\,\omega.
$$
So, Poynting theorem suggests to write down (denoting $\xi=ct$)
$$
L_{\mathbf{E}\times\mathbf{B}}\,\omega=
-\frac{\partial}{\partial\xi }\frac{\mathbf{E}^2+\mathbf{B}^2}{2}\,\omega,
$$
i.e.,
$$
\mathbf{B}.curl\,\mathbf{E}-\mathbf{E}.curl\,\mathbf{B}=-
\frac{\partial}{\partial\xi}\frac{\mathbf{E}^2+\mathbf{B}^2}{2}, \ \ \ \ \
\rightarrow\ \ \ \ \ \mathbf{B}.\left(
curl\,\mathbf{E}+\frac{\partial\mathbf{B}}{\partial\xi}\right)=
\mathbf{E}.\left(curl\,\mathbf{B}-\frac{\partial\mathbf{E}}{\partial\xi}\right).
$$
 These relations say: the difference of the two mutual
{\it local cross-helicities} $\mathbf{B}.curl\,\mathbf{E}$
and $\mathbf{E}.curl\,\mathbf{B}$ deforms the volume form
$\omega=dx\wedge dy\wedge dz$ by the $\xi$-derivative of the energy-density. Moreover, two of
the Maxwell equations are sufficient for this, and if the two fields are
nonzero only inside compact 3d-region $\mathbb{A}\subset\mathbb{R}^3$, for each
$t$, then
$$
L_{\mathbf{E}\times\mathbf{B}}\,\omega=div(\mathbf{E}\times\mathbf{B})\omega=
(\mathbf{B}.curl\,\mathbf{E}-\mathbf{E}.curl\,\mathbf{B})\omega=
\mathbf{d}(\beta\wedge\eta).
$$
Now the Stokes theorem with respect to $\mathbb{A}\subset\mathbb{R}^3$ leads to
zero of the integral
$\int_{\mathbf{R}^3}(L_{\mathbf{E}\times\mathbf{B}})\,\omega$, so,
$$
\frac{\partial}{\partial \xi}\int_{\mathbb{R}^3}
\frac{\mathbf{E}^2+\mathbf{B}^2}{2}\,\omega=0,
$$
i.e., the integral energy is conserved.

The relations considered suggest some connection with the concepts of {\it
absolute} and {\it relative} integral invariants of a vector field $X$ on a
manifold $M$ introduced and used by E.Cartan [23]: these are differential forms
$\alpha\in \Lambda(M)$ satisfying respectively the relations $i(X)\alpha=0,
i(X)\mathbf{d}\alpha=0$, leading to $L_{X}\alpha=0$, and just
$i(X)\mathbf{d}\alpha=0$. Our relations may be considered as corresponding
extensions: {\it a vector field} $\rightarrow$ {\it vector valued multivector
field}
and
{\it a differential form} $\rightarrow$  {\it vector valued
differential form}
making use of the mentioned in Sec.4.1 extension of the Lie derivative of a
differential form along multivector fields. The new moment in our extension is
that we consider vector valued multivectors along which vector valued forms to
be differentiated with respect to some bilinear map
$\varphi:V\times V\rightarrow W$, where $W$ is appropriately determined
vector space.

In general, we note that, the thriple $(V,W;\varphi)$ determines possible
interactions among the subsystems of the field object considered, which
subsystems are formally represented by the vector components of the multivector
(in our case $\mathbf{\bar\Omega}$) and the vector components of the
(multi)differential form (in our case $\mathbf{\Omega}$).

\vskip 0.3cm

\section{Conclusion}
Getting knowledge of the internal compatibility and external stability of a
physical object is being done by measuring the corresponding to these physical
appearances appropriate physical quantities. Such physical quantities may vary
in admissible, or not admissible degree: in the first case we talk about
admissible changes, and in the second case we talk about changes leading to
destruction of the object. Formally, this is ususlly checked by calculating the
flow of the formal image of the (sub)system considered through its
appropriately modeled change, as it is seen, e.g., in (8),(9),(11), i.e., by
means of finding corresponding {\it differential self flows of the subsystems},
e.g.,  $i_{\bar{F}}\mathbf{d}F$,
and {\it differential mutual flows among the subsystems}, e.g.,
$i_{\bar{F}}\mathbf{d}*F$. Since every
measuring process requires stress-energy-momentum transfering between the
object studied and the measuring system, the role of finding corresponding {\it
tensor} representatives of these change-objects and the corresponding flows is
of serious importance. Therefore, having adequate stress-energy-momentum for
the considered case, the clearly individualized tensor members of its
divergence represent qualitatively and quantitatively important aspects of the
{\it intrinsic interacting dynamical nature} of the object considered. This
view motivated the above given approach to find appropriate description of
electromagnetic field objects.

The existing knowledge about the structure and internal dynamics of free
electromagnetic field objects made us assume the notion for {\it two
partner-fields internal structure}, formally represented by $(F,*F)$ on
Minkowski space-time. Each of these two partner-fields is built of the two {\it
formal constituents} $(\mathbf{E},\mathbf{B})$, and each partner-field is able
to carry local stress-energy-momentum, allowing local "intercomunication"
between its two constituents during the local interaction with its
partner-field. The two subsystems carry equal local energy-momentum densities,
and realize local mutual energy exchange {\it without available interaction
energy}. Moreover, they strictly respect each other: the exchange is {\it
simultaneous} and in {\it equal quantities}, so, each of the two partner-fields
keeps its identity and recognizability. The corresponding internal dynamical
structure appropriately unifies translation and rotation through unique
space-time propagations as a whole with the fundamental velocity. All Maxwell
solutions are duly respected. The new nonlinear solutions, i.e., those
satisfying $\mathbf{d}F\neq 0, \mathbf{d}*F\neq 0$, are {\it time-stable, they
admit FINITE SPATIAL SUPPORT, and minimize the relation} $I_1^2+I_2^2 \geq 0$.
It deserves noting here that the obtained relation $I_1^2+I_2^2=0$ for the
nonlinear solutions is equivalent to
$I_1=\frac12F_{\mu\nu}F^{\mu\nu}=\mathbf{B}^2-\mathbf{E}^2=0, \ \
I_2=\frac12F_{\mu\nu}(*F)^{\mu\nu}=2\mathbf{E}.\mathbf{B}=0$ (for details see
[21]).
\vskip 0.4cm
The admitted solutions with spatially finite support are of {\it photon-like
nature}:
 \vskip 0.3cm
-they are time-stable,
 \vskip 0.1cm
-they demonstrate intrinsically compatible translational-rotational
dynamical structure,
\vskip 0.3cm
-they propagate translationally as a whole with the velocity of light,
 \vskip 0.1cm
-they carry finite energy-momentum and intrinsically
determined integral characteristic $\mathfrak{h}$ of action nature through
naturally available appropriate scale factor $\mathcal{L}_o=const$
[21,pp.233] carrying physical dimension of length,
 \vskip 0.1cm
-their integral energy $E$ satisfies relation of
the form identical to the Planck formula $E.T=\mathfrak{h}$ [21,pp.230-231].

\vskip 0.3cm
Some of these nonlinear solutions of (12) look like:

\begin{center}
\begin{figure}[ht!]
\centerline{
{\mbox{\psfig{figure=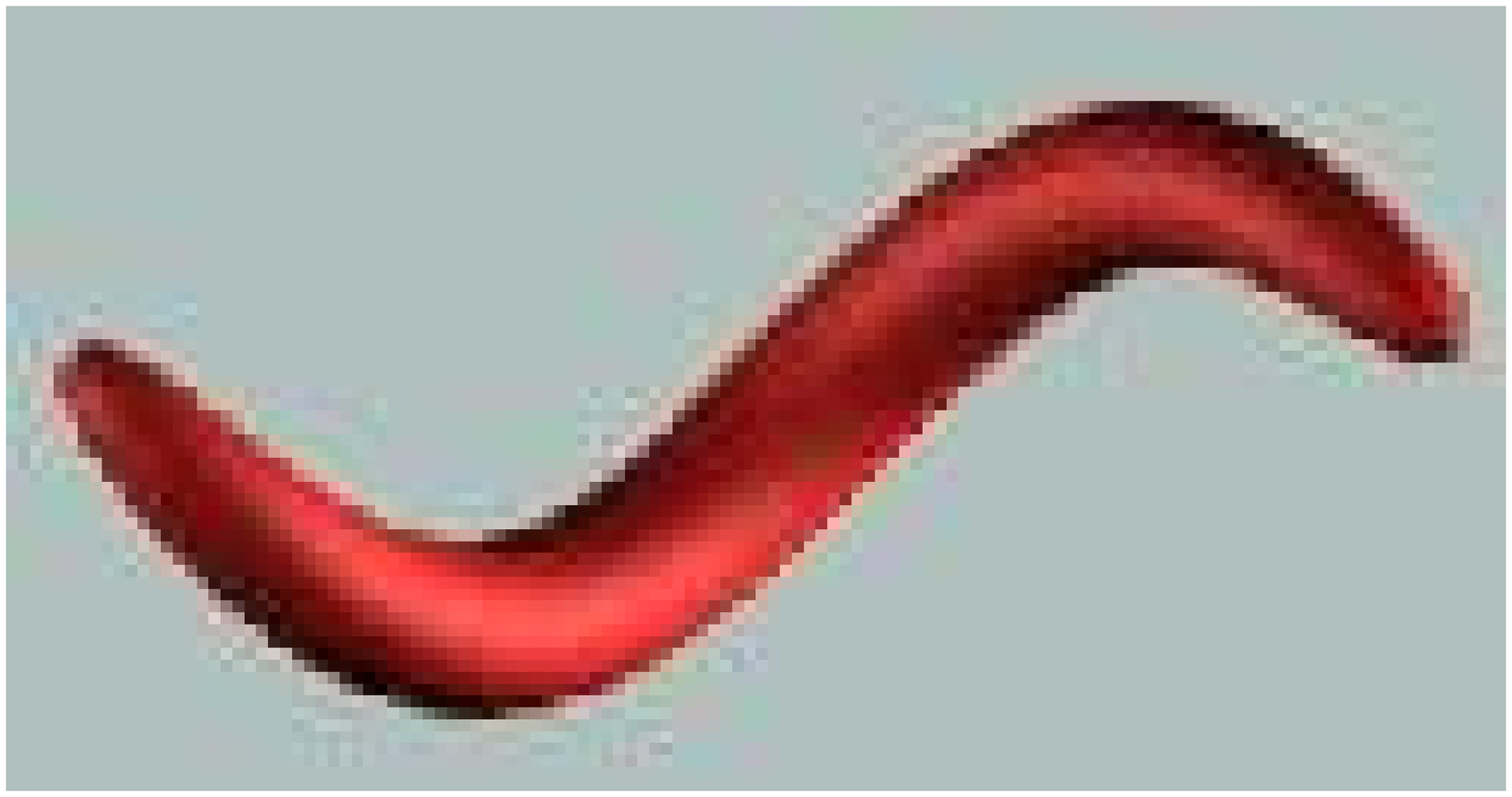,height=1.8cm,width=3.5cm}}
\mbox{\psfig{figure=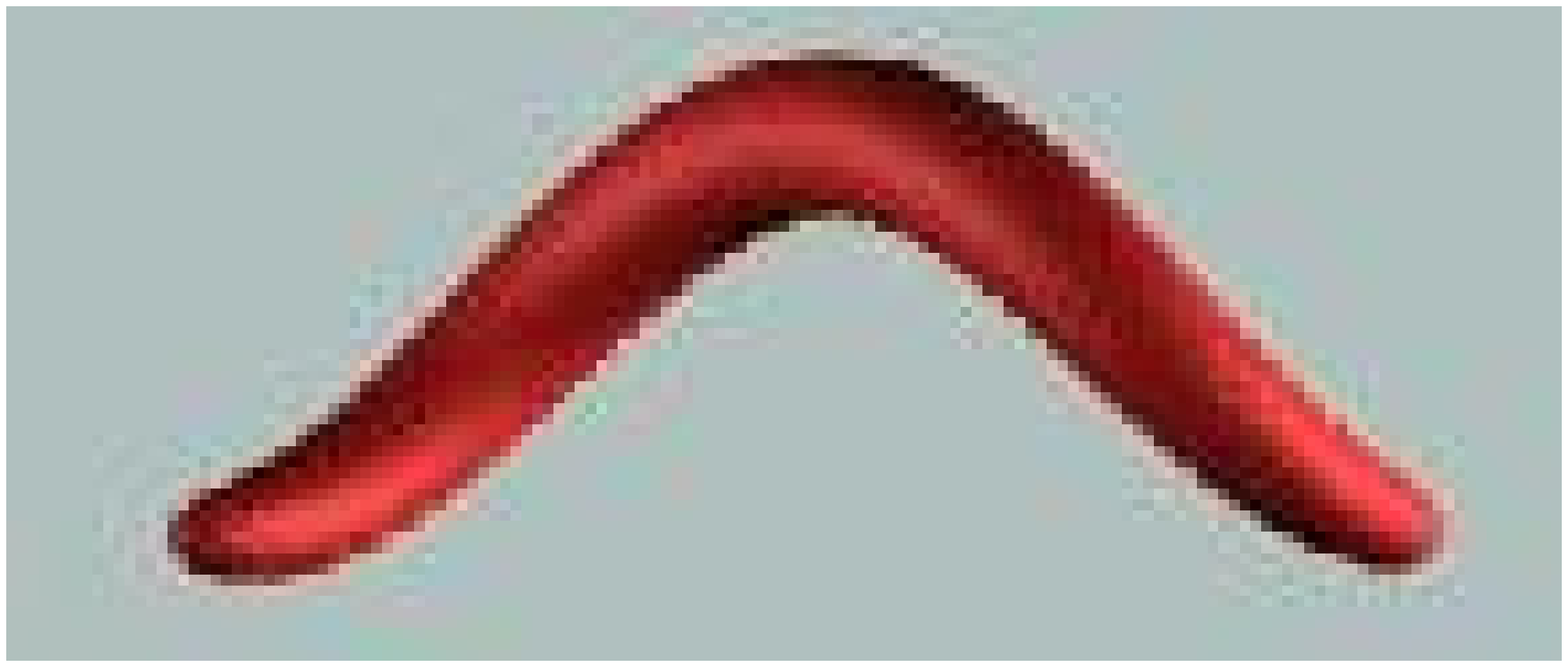,height=1.8cm,width=4.2cm}}
\mbox{\psfig{figure=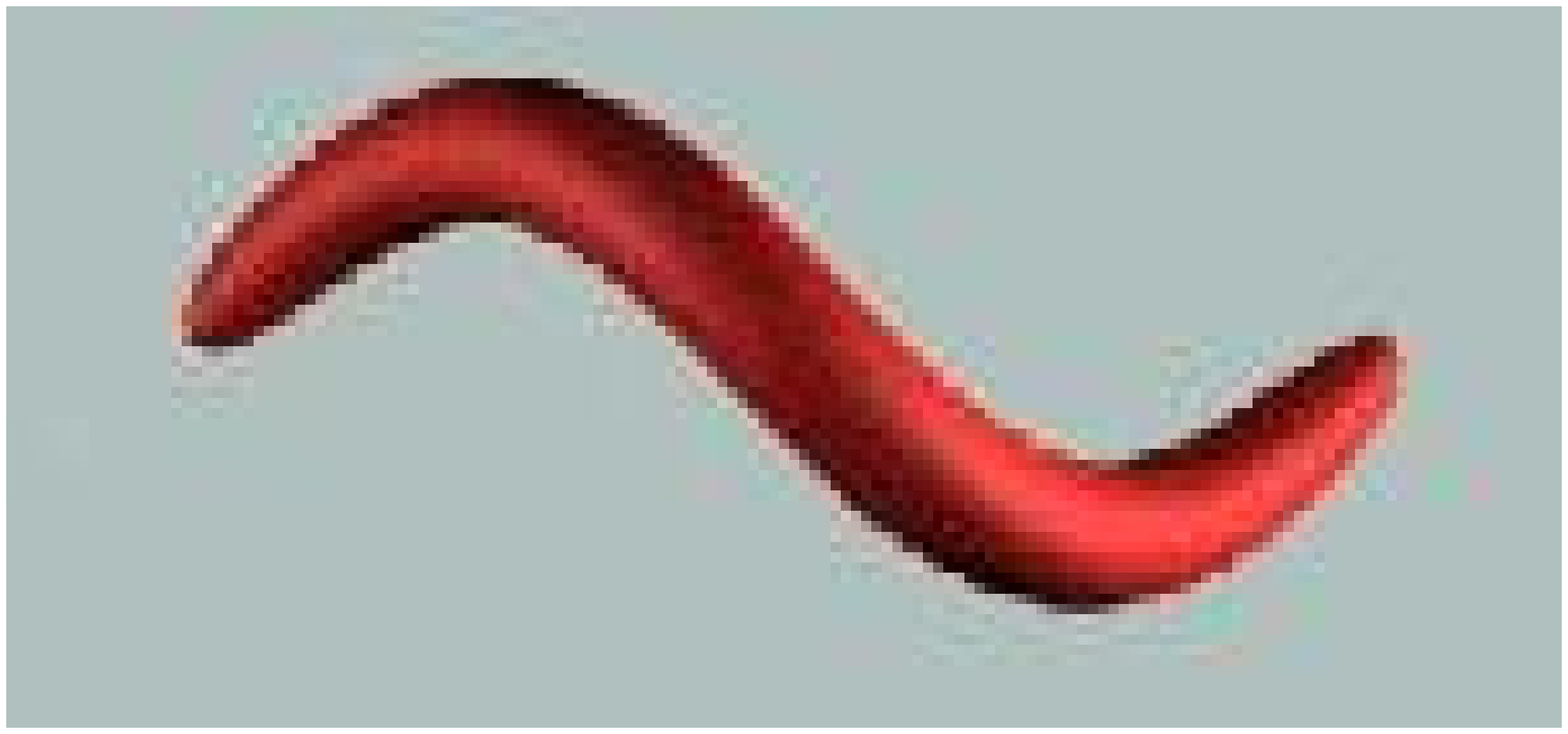,height=1.8cm,width=4.2cm}}}}
\caption{Theoretical example with clock-wise rotation, translation: left to
right}
\end{figure} \end{center}
\begin{center} \begin{figure}[ht!]
\centerline{ {\mbox{\psfig{figure=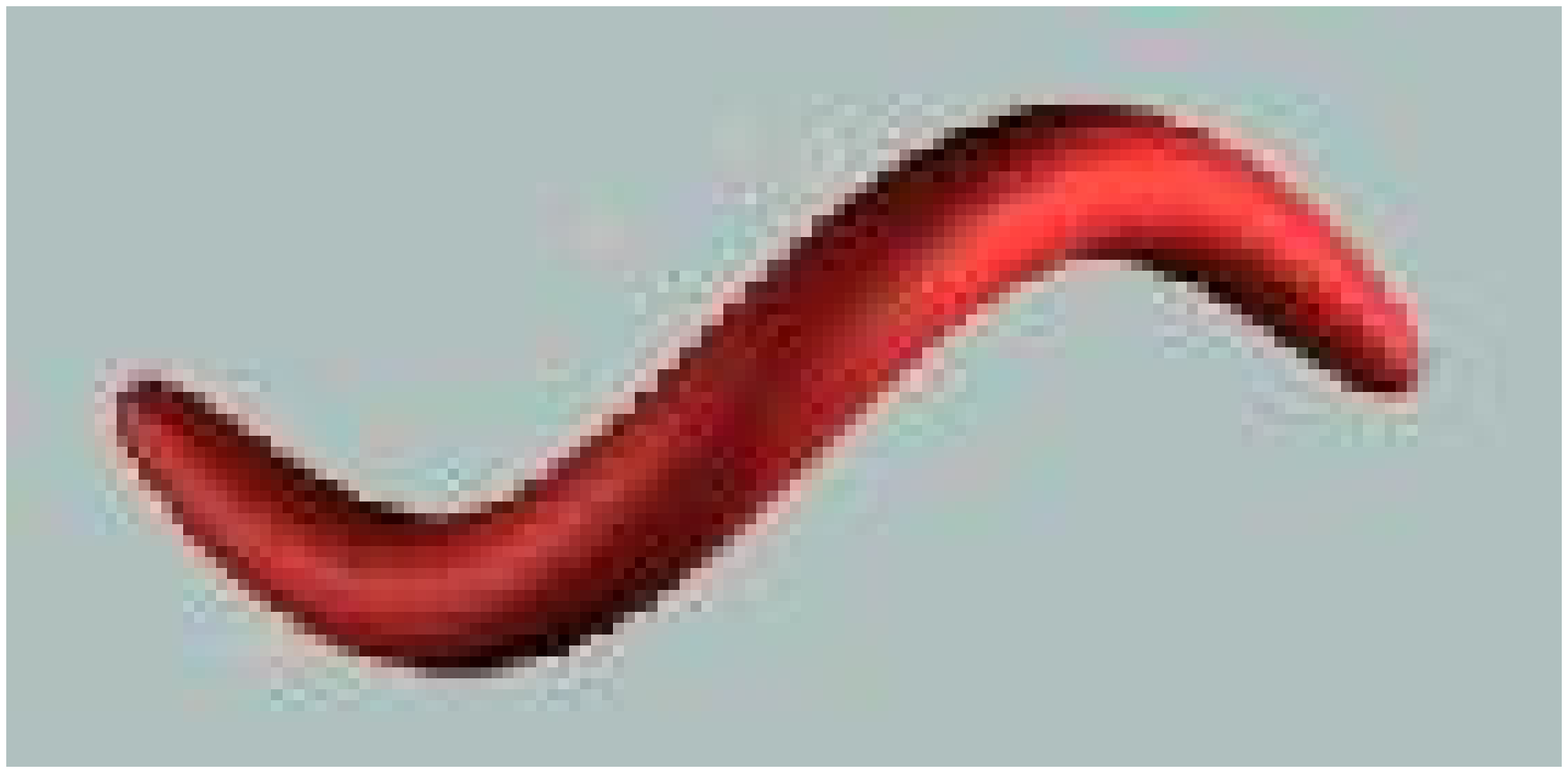,height=1.8cm,width=3.5cm}}
\mbox{\psfig{figure=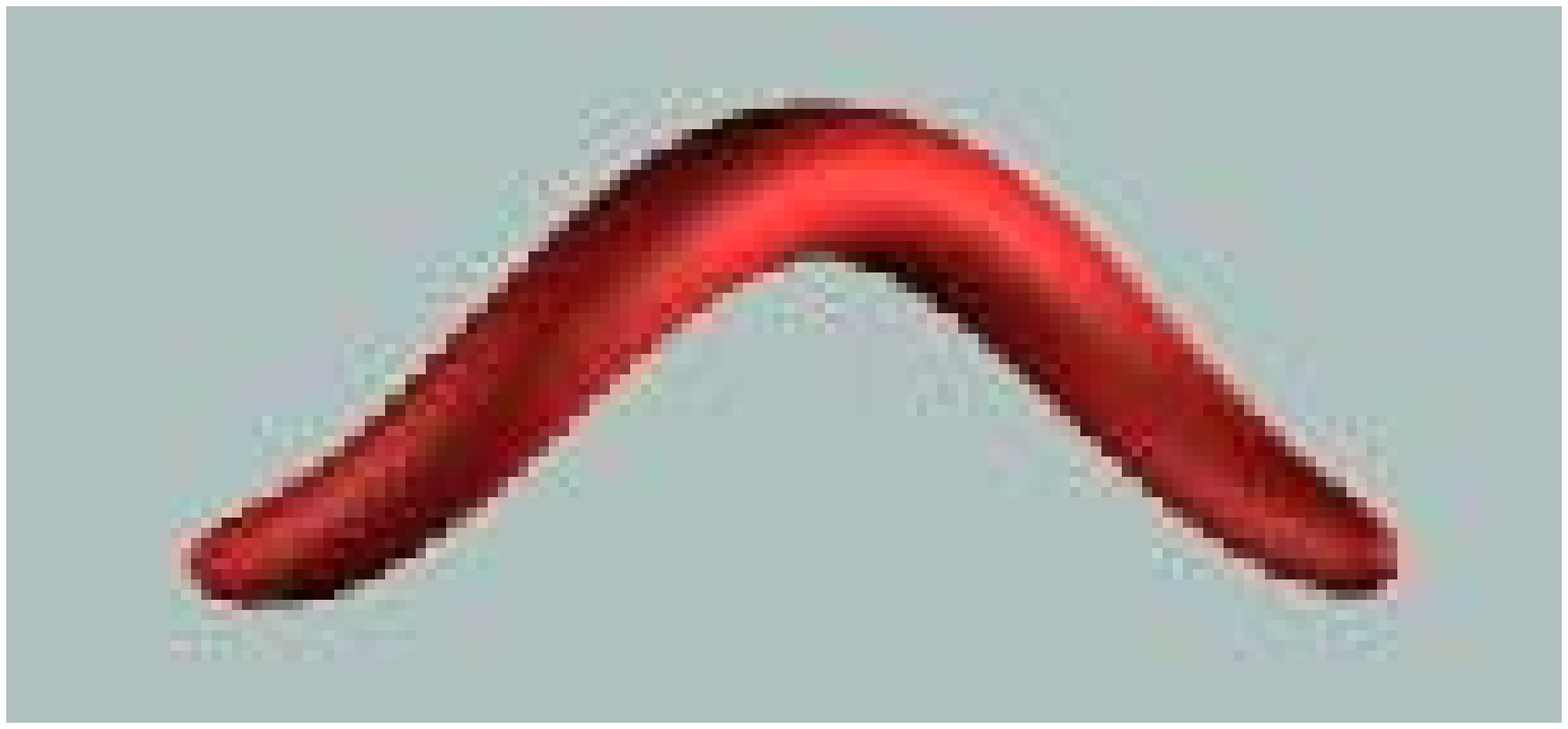,height=1.8cm,width=4.2cm}}
\mbox{\psfig{figure=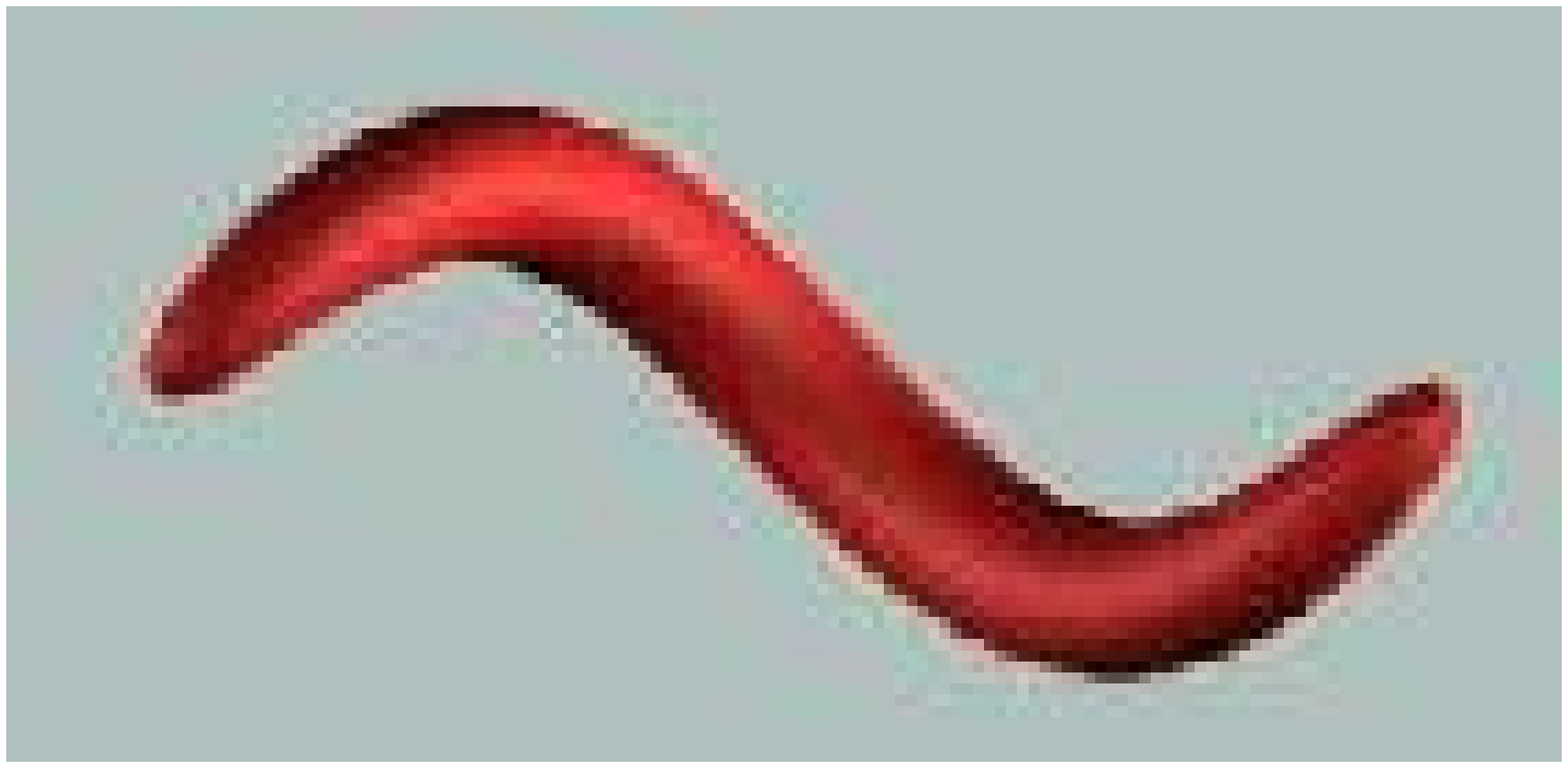,height=1.8cm,width=4.2cm}}}}
\caption{Theoretical example with anticlock-wise rotation, translation: left to
right}
 \end{figure}
\end{center}

 As we mentioned above, the straight line size along translational
propagation of each of these finite helical-like objects is
$2\pi\mathcal{L}_o=const$, $T=\frac{2\pi\mathcal{L}_o}{c}$, so, $\mathfrak{h}$
is an integral Lorentz invariant action characteristic of any solution of this
class, meaning: there is specific propagational action demonstrated during the
intrinsically defined time period $\frac{2\pi\mathcal{L}_o}{c}$, where $c$ is
the invariant speed of translational propagation as a whole.

\newpage
{\bf References}
\vskip 0.3cm

[1]. {\bf M. Born, L. Infeld}, {\it Nature}, {\bf 132}, 970 (1932)

[2]. {\bf M. Born, L.Infeld}, {\it Proc.Roy.Soc.}, {\bf A 144}, 425 (1934)

[3]. {\bf W. Heisenberg, H. Euler}, {\it Zeit.Phys.}, {\bf 98}, 714 (1936)

[4]. {\bf M. Born}, {\it Ann. Inst. Henri Poincare}, {\bf 7}, 155-265 (1937).

[5]. {\bf J. Schwinger}, {\it Phys.Rev}. ,{\bf 82}, 664 (1951).

[6]. {\bf H. Schiff}, {\it Proc.Roy.Soc.} {\bf A 269}, 277 (1962).

[7]. {\bf J. Plebanski}, {\it Lectures on Nonlinear Electrodynamics}, NORDITA,
Copenhagen, 1970.

[8]. {\bf G. Boillat}, {\it Nonlinear Electrodynamics: Lagrangians and
Equations of Motion}, \newline J.Math.Phys. {\bf 11}, 941 (1970).

[9]. {\bf B. Lehnert, S. Roy}, {\it Extended Electromagnetic Theory}, World
Scientific, 1998.

[10]. {\bf D.A. Delphenich}, {\it Nonlinear Electrodynamics and QED},
arXiv:hep-th/0309108, (good review article).

[11]. {\bf B. Lehnert}, {\it A Revised Electromagnetic Theory with Fundamental
Applications}, Swedish Physic Arhive, 2008.

[12]. {\bf D. Funaro}, {\it Electromagnetsm and the Structure of Matter},
Worldscientific, 2008; also: {\it From photons to atoms}, arXiv: gen-ph/1206.3110
(2012).

[13]. {\bf E. Schrodinger}, {\it Contribution to Born's new theory of
electromagnetic feld}, Proc. Roy. Soc. Lond. {\bf A 150}, 465 (1935).

[14]. {\bf G. Gibbons, D. Rasheed}, {\it Electric-magnetic duality rotations in
non-linear electrodynamics}, Nucl. Phys. {\bf B 454} 185 (1995) hep-th/9506035.

[15] {\bf R. Kerner, A.L. Barbosa, D.V. Gal'tsov}, {\it Topics in Born-Infeld
Electrodynamics}, arXiv: hep-th/0108026 v2

[16]. {\bf J. Marsden, A. Tromba}, {\it Vector Calculus}, fifth edition, W.H.
Freeman and Company, 2003.

[17]. {\bf W.M. Tulczyjew}, {\it The Graded Lie Algebra of Multivector Fields
and the Generalized Lie Derivative of Forms}, Bull. Acad. Pol. Sci. SMAP 22
(1974) 937-942; {\it The Poisson Bracket for Poisson Forms in Multisymplectic
Field Theory}, arXiv: math-ph/0202043v1

[18]. {\bf S.G.Donev}, {\it A particular nonlinear generalization of Maxwell
equations admitting spatially localized wave solutions},
Compt.Rend.Bulg.Acad.Sci., vol.34, No.4 (1986).

[19]. {\bf S. Donev, M. Tashkova}, {\it Energy-momentum directed
nonlinearization of Maxwell's pure field equations}, Proc.R.Soc.Lond. A ,
1993, {\bf 443}, 301-312.

[20]. {\bf S. Donev, M. Tashkova},
{\it Energy-Momentum Directed Nonlinearization of Maxwell's  Equations in the
Case of a Continuous Medium} /Donev, S., Tashkova, M./,   Proc.R.Soc. Lond.A
{\bf 450}, 281 (1995)

[21]. {\bf S. Donev, M. Tashkova}, {\it Geometric View on Photon-like Objects},
 LAMBERT Academic Publishing, 2014 (also: arXiv,math-ph, 1210.8323v2)

[22] 1. {\bf D.Reed}, {\it Foundational Electrodynamics and Beltrami Vector
Fields}, in Advanced Electromagnetism: Foundations, Theory, Applications, D.
Grimes, T.W. Barrett (eds), World Scientific, Singapore, 1995 ; \
2. {\bf E. Beltrami},
{\it Considerations on Hydrodynamics}, Rendiconti del Reale Instituto
Lombardo…Series II, vol. 22, 1889(trans. By G. Filliponi, Int. J. Fusion
Energy, 3(3), pp. 51-57, 1985), \
3. {\bf O. Bjorgum},
{\it On Beltrami Vector Fields and Flows: A comparative study of some basic
types of vector fields} , Universitetet i Bergen ; \ \ 4.
{\bf O. Bjorgum, T. Godal}, {\it On Beltrami Vector
Fields and Flows(Part II)}, Universitet I Bergen Arbok, 1952.

[23]. {\bf E.Cartan}, {\it Lecons sur les invariants integraux}.
Cours professe a la Faculte des sciences de Paris, 1920-1921.

 \end{document}